\PassOptionsToPackage{hyphens}{url}
\documentclass[11pt]{article} 
\usepackage[a4paper,left=2.5cm, right=2.5cm, top=2.5cm, bottom=2.5cm]{geometry}
\usepackage{amsmath,amssymb}
\usepackage{setspace}
\usepackage{iftex}
\ifPDFTeX
  \usepackage[T1]{fontenc}
  \usepackage[utf8]{inputenc}
  \usepackage{textcomp} 
\else 
  \usepackage{unicode-math} 
  \defaultfontfeatures{Scale=MatchLowercase}
  \defaultfontfeatures[\rmfamily]{Ligatures=TeX,Scale=1}
\fi
\usepackage{lmodern}

\IfFileExists{upquote.sty}{\usepackage{upquote}}{}
\IfFileExists{microtype.sty}{
  \usepackage[]{microtype}
  \UseMicrotypeSet[protrusion]{basicmath} 
}{}
\makeatletter
\@ifundefined{KOMAClassName}{
  \IfFileExists{parskip.sty}{%
    \usepackage{parskip}
  }{
    \setlength{\parindent}{0pt}
    \setlength{\parskip}{6pt plus 2pt minus 1pt}}
}{
  \KOMAoptions{parskip=half}}
\makeatother
\usepackage{xcolor}
\providecommand{\tightlist}{%
  \setlength{\itemsep}{0pt}\setlength{\parskip}{0pt}}
\makeatletter
 \let\@cite@ofmt\@firstofone
 \def\@biblabel#1{}
 \def\@cite#1#2{{#1\if@tempswa , #2\fi}}
\makeatother
\newlength{\cslhangindent}
\newlength{\csllabelwidth}
 {\begin{list}{}{%
  \setlength{\itemindent}{0pt}
  \setlength{\leftmargin}{0pt}
  \setlength{\parsep}{0pt}
  \ifodd #1
   \setlength{\leftmargin}{\cslhangindent}
   \setlength{\itemindent}{-1\cslhangindent}
  \fi
  \setlength{\itemsep}{#2\baselineskip}}}
 {\end{list}}
\usepackage{calc}

\usepackage{booktabs}
\usepackage{longtable}
\usepackage{array}
\usepackage{multirow}
\usepackage{wrapfig}
\usepackage{colortbl}
\usepackage{pdflscape}
\usepackage{tabu}
\usepackage{threeparttable}
\usepackage{threeparttablex}
\usepackage[normalem]{ulem}
\usepackage{makecell}
\usepackage{xcolor}
\usepackage{svg}
\usepackage{pdfpages}
\usepackage{titlesec}
\titlespacing{\title}{0pt}{\parskip}{-\parskip}
\usepackage{floatrow}
\usepackage{algorithm}
\usepackage[noend]{algpseudocode}
\usepackage{soul}
\usepackage{ragged2e}
\usepackage{float}
\usepackage[
    backend=biber,
    uniquename=false,
    style=authoryear,
    sorting=nyt,
    maxcitenames=2
]{biblatex}
\ifLuaTeX
  \usepackage{selnolig}  
\fi
\usepackage{bookmark}
\IfFileExists{xurl.sty}{\usepackage{xurl}}{} 
\hypersetup{
  hidelinks,
  pdfcreator={LaTeX via pandoc}}

\usepackage{xcolor}
\usepackage{fancyhdr}

\usepackage{makecell}
\usepackage{amssymb}
\usepackage{csquotes}
\usepackage{amsmath}
\usepackage{graphicx}
\usepackage{tabularx}
\usepackage{enumitem}
\usepackage{pdfpages}
\usepackage{authblk}

\definecolor{WNEcolor}{HTML}{C00000}

\newlist{questions}{itemize}{1}
\setlist[questions]{label=\textbf{Q:}}

\title{A novel approach to trading strategy parameter
optimization, using double out-of-sample data and
walk-forward techniques\thanks{The associated GitHub repository covering all calculations presented in this paper can be found at: \url{https://github.com/tmr-crypto/wf_optim_crypto_analysis}}}
\date{\vspace{-5ex}}

\author[1]{Tomasz Mroziewicz}
\author[2]{Robert Ślepaczuk}
\affil[1]{\small University of Warsaw, Faculty of Economic Sciences, Ul. Długa 44/50, 00-241 Warsaw, Poland, ORCID: https://orcid.org/0009-0003-6540-6554, email: tomasz.mroziewicz2@gmail.com}
\affil[2]{\small University of Warsaw, Faculty of Economic Sciences, Department of Quantitative Finance and Machine Learning, Quantitative Finance Research Group, Ul. Długa 44/50, 00-241 Warsaw, Poland, ORCID: https://orcid.org/0000-0001-5227-2014, Corresponding author: rslepaczuk@wne.uw.edu.pl }

\begin{document}

\maketitle
\begin{abstract}
    This study introduces a novel approach to walk-forward optimization by parameterizing the lengths of training and testing windows. We demonstrate that the performance of a trading strategy using the Exponential Moving Average (EMA) evaluated within a walk-forward procedure
based on the Robust Sharpe Ratio is highly dependent on the chosen window size. We
investigated the strategy on intraday Bitcoin data at six frequencies (1 minute to 60 minutes)
using 81 combinations of walk-forward window lengths (1 day to 28 days) over a 19-month training period. The two best-performing parameter sets from the training data were applied to a 21-month out-of-sample testing period to ensure data independence. The strategy was only
executed once during the testing period. To further validate the framework, strategy parameters
estimated on Bitcoin were applied to Binance Coin and Ethereum. Our results suggest the robustness of our custom approach. In the training period for Bitcoin, all combinations of walk-forward windows outperformed a Buy-and-Hold strategy. During the testing period, the strategy performed similarly to Buy-and-Hold but with lower drawdown and a higher Information Ratio. Similar results were observed for Binance Coin and Ethereum. The real strength was demonstrated
when a portfolio combining Buy-and-Hold with our strategies outperformed all individual
strategies and Buy-and-Hold alone, achieving the highest overall performance and a 50\% reduction in drawdown. A conservative fee of 0.1\% per transaction was included in all calculations. A cost sensitivity analysis was performed as a sanity check, revealing that the strategy's break-even point was around 0.4\% per transaction. This research highlights the importance of optimizing walk-forward window lengths and emphasizing the value of single-time out-of-sample testing for reliable strategy evaluation.
\end{abstract}

{
\setcounter{tocdepth}{2}

}
\setstretch{1.5}

\newpage
\hypertarget{introduction}{%
\section{Introduction}\label{introduction}}

The development of trading strategies in stock and cryptocurrency markets has been extensively explored through technical analysis in prior research. However, these studies frequently overlook critical elements such as transaction fees, slippage, and other trading costs. Additionally, a common but often unreported issue involves repeated optimizations on data designated as out-of-sample, or the complete omission of an out-of-sample evaluation phase. Such practices foster unrealistic performance expectations, leading to substantial underperformance when strategies are implemented in live trading environments.
\newline
This study addresses key challenges in trading strategy development by providing a practical framework for constructing and assessing trading strategies. The framework incorporates tools for profitability estimation, statistical significance evaluation, and overfitting risk mitigation. Central to this investigation is the integration of the Exponential Moving Averages (EMA) Crossover—a momentum-based technical analysis strategy—with the walk-forward optimization procedure, which serves as the foundational structure of our approach.
\newline
The study investigates the following research questions: \newline
RH1: Walk-forward optimization, a method widely used in this work, divides the research period into training/testing subperiods instead of optimizing parameters over the entire dataset, which risks overfitting. For each training subperiod, optimal parameters are identified and evaluated out-of-sample in the corresponding testing subperiod. Out-of-sample results across all testing periods are aggregated to assess overall performance. This study examines the effects of varying training and testing window lengths on risk-adjusted returns, aiming to identify any dependence between window length and investor outcomes. It explores whether shorter or longer windows influence results, contrasting with the Efficient Market Hypothesis \parencite{fama1965behavior}, which suggests market efficiency precludes significant performance differences from window length variations.\newline
RH2: How do the choice of data frequency and costs impact strategy results?
This research utilizes intraday high-frequency data, with frequencies ranging from 1 to 60 minutes evaluated. The paper assesses the influence of costs and time frequency on risk-adjusted returns.
\newline
RH3: Is the tested strategy better than random?
To evaluate statistical significance, the bootstrap methodology is applied to determine if the optimal strategy outperforms randomly constructed alternatives generated during simulations.
\newline
Walk-forward optimization, an established technique in time series analysis, is employed to avoid overfitting in parameter estimation. This study extends beyond conventional fixed window lengths by examining the effects of varying training and testing window lengths on strategy performance, potentially enhancing strategy efficacy and contributing to a broader theoretical understanding of walk-forward optimization. Data utilization is approached innovatively by partitioning available data into a global training period for parameter optimization and an unseen period for execution, thereby reducing data mining bias. This differs from typical practices where strategies are iteratively tested on purported out-of-sample data, compromising their integrity. The analysis incorporates intraday cryptocurrency data, including Bitcoin, Ethereum, and Binance Coin, with parameters derived from Bitcoin training applied across assets to test adaptability.
\newline
A novel aspect of this work is the systematic investigation of training and testing window lengths in walk-forward optimization, moving away from arbitrary fixed lengths to uncover their impact on performance. Results indicate that the EMA Crossover strategy combined with walk-forward optimization reduces volatility and drawdown relative to Buy-and-Hold approaches. When integrated into portfolios with Buy-and-Hold, it enhances all performance metrics. The strategy's parameters, optimized on Bitcoin, yield superior results when applied to Ethereum and Binance Coin in the testing period. Furthermore, constructing portfolios blending passive Buy-and-Hold with the active EMA-walk-forward strategy reveals untapped diversification benefits, achieving optimal risk-adjusted measures and drawdown reductions compared to standalone Buy-and-Hold or walk-forward strategies. Overall, the framework demonstrates adaptability to evolving market regimes, with the strategy delivering positive returns for nearly two years on intraday cryptocurrency data without retraining. This persistence prompts further inquiry into potential improvements via periodic retraining during testing.
\newline
The paper is organized as follows: Section 2 presents a review of relevant academic literature. Section 3 details the methodology employed. Subsequent sections report the results, including analyses of cost sensitivity and seasonality in strategy returns. Finally, results on unseen data are discussed, followed by conclusions.

\hypertarget{literature-review}{%
\section{Literature Review}\label{literature-review}}

Technical analysis remains underrecognized in academia, likely due to its simplicity and lack of alignment with financial ratios or economic indicators. Despite limited scientific explanations, a Group of Thirty (1985) survey found that 97

\textcite{Park2007} reviewed 92 studies on technical analysis, showing 58 profitable, 24 negative, and 10 mixed results. Pre-1980s strategies were profitable in futures and foreign exchange but not in stocks. A recent review by \textcite{Nti2020} of 122 papers on technical and fundamental analysis reported 87

\textcite{Hsu2010} noted that ETFs may reduce technical strategy profitability, yet inefficiencies persist in immature markets, attracting arbitrageurs and enhancing efficiency. Challenges include data limitations and assumptions of closing-price executions. \textcite{zielonka2004technical} found investors overestimate technical signals due to cognitive biases like confirmation bias, favoring aligned beliefs; simple rules like moving average crossover are often neglected.

\textcite{Nefti1991NaiveTR} described moving average crossover as statistically well-defined, unlike most technical rules. \textcite{FARIASNAZARIO2017115} identified it as the most popular method (44 of 80 studies), surpassing neural networks or genetic algorithms (15 each). Hybrid approaches include \textcite{Ayala2021}, combining TEMA and MACD with SVMs and Random Forests for positive results on stock indices, excluding costs. \textcite{Karasu2020} integrated EMA, SMA, Kaufman's adaptive moving average, MOPSO, and RBFSVR to improve crude oil price forecasting over traditional methods.

\textcite{brock1992simple} tested moving average crossover, yielding higher returns and lower volatility post-buy signals and negative returns with higher volatility post-sell, outperforming AR(1), GARCH, and random walk models. \textcite{Zakamulin2020} compared momentum and moving average crossover, finding similarities in strong trends; under random walk, they equate to coin flips, but with trends or mean reversion, they exceed 50

As a recent phenomenon, cryptocurrencies have drawn growing academic interest with mixed results. \textcite{Brown2019} applied moving average crossover to 40 cryptocurrencies, outperforming Buy-and-Hold in only 10; 10 lacked data, and Buy-and-Hold won in the remaining 20.

Walk-forward analysis divides time series into sequential in-sample and out-of-sample periods, optimizing parameters in-sample and evaluating out-of-sample. \textcite{Pardo2012a} argued it reduces overfitting, boosts reliability, and assesses market adaptation, though it requires substantial data and computation. \textcite{LopezdePrado2019} favored Monte Carlo with synthetic data over walk-forward, as the latter assumes consistent data processes and ignores stochasticity. \textcite{LeBaron1998_bootstrap} critiqued train/validation/test splits for overestimating performance, using bootstrapping to show variability from splits, not parameters.

\textcite{Karathanasopoulos2016} used 300- and 400-day sliding windows to predict crude oil-refined product spreads, finding 300-day windows with particle swarm optimization and radial basis function neural networks slightly superior in risk-adjusted returns. In cryptocurrencies, \textcite{AHMED2020101495} applied moving average crossover to Dash (2016-2018), yielding 14-18\% excess over Buy-and-Hold, but not for a 10-coin privacy portfolio. \textcite{Wen2022} identified intraday predictability, momentum, and reversal in cryptocurrencies, impacted by events like FOMC meetings and COVID-19; strategies outperformed Buy-and-Hold, excluding costs.

\textcite{AlonsoMonsalve2020} used neural networks with one-minute data and technical indicators to predict cryptocurrency returns, omitting costs. \textcite{Wei2018} found no illiquidity premium in 456 cryptocurrencies, with illiquid ones showing antipersistence (low Hurst exponents) and deviations from randomness, unlike efficient liquid markets. \textcite{ANGHEL2021} applied White's Reality Check to technical and machine learning strategies, finding rare significant returns after data snooping and frictions; machine learning underperformed technical analysis due to higher transactions, with no added benefits over traditional markets. \textcite{Kosc2019} examined momentum and contrarian effects in 100 large-cap cryptocurrencies, with short-term contrarian outperforming momentum and benchmarks, offering diversification over the S\&P 500.

Limitations persist: \textcite{FARIASNAZARIO2017115} noted only 37\% of studies adjust for risk. \textcite{Park2007} highlighted biases like data snooping, rule selection, and cost estimation difficulties. \textcite{Efron1977} popularized bootstrapping for i.i.d. data, but financial series violate assumptions via autocorrelation and heteroscedasticity, per \textcite{Ruiz2002}; block bootstrapping by \textcite{Politis1992} addresses this.

Presented research provides insights into technical analysis, though academia lacks consensus on its effectiveness or foundations. This study contributes additional questions to the discussion rather than definitive answers.

\hypertarget{data}{%
\section{Data}\label{data}}

\hypertarget{description-of-data}{%
\subsection{Description of data}\label{description-of-data}}

The analyzed time series represent cryptocurrency prices. The data was acquired from Kaggle.com, which hosted a predictive modeling challenge called G-Forecast in collaboration with a commercial partner. The contest data sampled the prices of 10 different cryptocurrencies at one-minute intervals, although only a subset was ultimately used in this
research.

The data encompasses two distinct periods. Notably, the Unseen Data Period is slightly longer than the Global Training Data Period:

\begin{itemize}
\tightlist
\item
  Global Training Data Period: February 8, 2018, to September 1, 2019
\item
  Unseen Data Period: November 7, 2019, to August 22, 2021.
\end{itemize}

The gap between the end of the Global Training Data Period and the start of the Unseen Data Period arises from a walk-forward adjustment to ensure both
periods begin and end at the same time, despite potential differences in
data frequency.

\begin{figure}[H]
\includegraphics{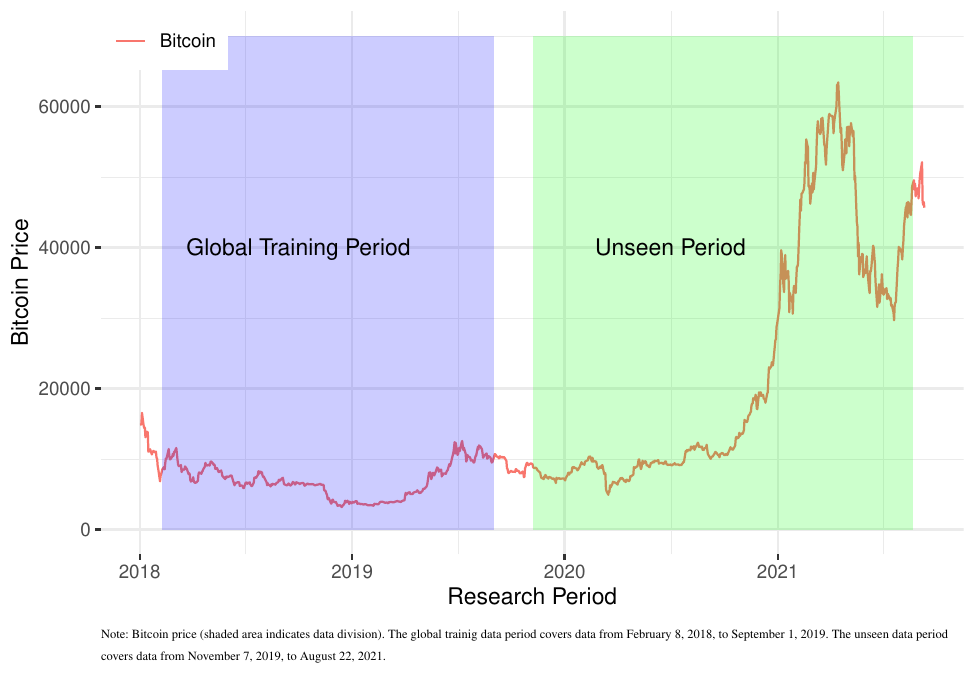} \caption{\label{bitcoin_price} Price of Bitcoin with marked Global Training/Unseen Period}\label{fig:bitcoin_price }
\end{figure}

\hypertarget{description-of-the-global-training-data-period}{%
\subsection{Description of the Global Training Data
Period}\label{description-of-the-global-training-data-period}}

\hypertarget{basic-statistic-properties}{%
\subsection{Basic statistic
properties}\label{basic-statistic-properties}}

Table \ref{desc_stats} explores the basic statistical properties of the
Bitcoin price time series across various sampling intervals.

Key findings:

\begin{itemize}
\tightlist
\item
  While high kurtosis is observed consistently, interestingly, it
  exhibits a decreasing trend with increasing sampling intervals.
\item
  This pattern is disrupted at the 5-minute interval, where kurtosis
  remains higher compared to the 1-minute interval. Further investigation is needed to understand the behavior at the 5-minute sampling. High kurtosis indicates a higher probability of large price movement and a leptokurtic feature of time series; findings align with previous studies, such as the work by \textcite{AllaA.Petukhina2021}, where the kurtosis of intraday data was estimated. It is worth noting that
  the mentioned research has documented kurtosis for Bitcoin as high as 49 for 5-minute data, which is much higher than presented below.
\item
  Across all sampling intervals, the data exhibit positive skew,
  meaning the right tail extends further than the left. This indicates a
  higher probability of large positive price movements compared to large
  negative ones.
\item
  Additionally, the maximum values observed in each sampling interval
  are consistently larger in absolute terms for positive deviations than
  for negative deviations, further confirming the asymmetry of the
  distribution.
\end{itemize}

\begin{table}[H]

\caption{\label{tab:descripitve_stats}\label{desc_stats} Descriptive statistics of the global training data for Bitcoin (BTC)}
\centering
\fontsize{10}{12}\selectfont
\begin{threeparttable}
\begin{tabular}{rrrrrrrrrr}
\toprule
\makecell[r]{Freq\\Minutes} & Mean & SD & Median & Min & Max & Range & Skew & Kurtosis & \makecell[r]{JB\\P-Value}\\
\midrule
\cellcolor{gray!6}{1} & \cellcolor{gray!6}{0} & \cellcolor{gray!6}{0.0013} & \cellcolor{gray!6}{0.0000} & \cellcolor{gray!6}{-0.0465} & \cellcolor{gray!6}{0.0315} & \cellcolor{gray!6}{0.0779} & \cellcolor{gray!6}{-0.0769} & \cellcolor{gray!6}{18.13} & \cellcolor{gray!6}{0}\\
5 & 0 & 0.0027 & 0.0000 & -0.0792 & 0.0589 & 0.1382 & 0.0102 & 31.97 & 0\\
\cellcolor{gray!6}{10} & \cellcolor{gray!6}{0} & \cellcolor{gray!6}{0.0037} & \cellcolor{gray!6}{0.0000} & \cellcolor{gray!6}{-0.0760} & \cellcolor{gray!6}{0.0610} & \cellcolor{gray!6}{0.1369} & \cellcolor{gray!6}{-0.0324} & \cellcolor{gray!6}{26.16} & \cellcolor{gray!6}{0}\\
15 & 0 & 0.0045 & 0.0000 & -0.1364 & 0.0770 & 0.2134 & -0.3408 & 37.87 & 0\\
\cellcolor{gray!6}{30} & \cellcolor{gray!6}{0} & \cellcolor{gray!6}{0.0064} & \cellcolor{gray!6}{0.0000} & \cellcolor{gray!6}{-0.0878} & \cellcolor{gray!6}{0.1033} & \cellcolor{gray!6}{0.1911} & \cellcolor{gray!6}{0.0775} & \cellcolor{gray!6}{22.63} & \cellcolor{gray!6}{0}\\
\addlinespace
60 & 0 & 0.0089 & 0.0001 & -0.0925 & 0.1081 & 0.2006 & 0.1598 & 16.08 & 0\\
\bottomrule
\end{tabular}
\begin{tablenotes}[para]
\item \textit{\tiny{ Note: }} 
\item \tiny{ Descriptive statistics of the global training data for Bitcoin (BTC).  JB P-Value - Jarque–Bera test p-value. The global training data period covers data from February 8, 2018, to September 1, 2019. All statistics are presented on a non-annualized basis. }
\end{tablenotes}
\end{threeparttable}
\end{table}

\hypertarget{methodology}{%
\section{Methodology}\label{methodology}}

The study employed various methodologies to minimize the influence of
cognitive biases in strategy results. To decrease the risk of data
mining bias, during the research architecture design phase, the data was
divided into two main periods: the Global Training Data Period and the
Unseen Data Period. All strategy parameter optimization was performed on
the Global Training Data Period, while the Unseen Data Period was used
only once to calculate the performance for the final best parameter
combination. Using the Unseen Data Period only once ensures that the
results reflect the strategy's genuine performance on truly unseen data
and are less susceptible to data mining bias.

A novel approach to strategy optimization was used. Instead of solely
finding the best Moving Average crossover parameters, an additional
level of optimization involved finding the optimal lengths of
walk-forward training/testing periods. In typical studies, the length of
walk-forward steps is chosen arbitrarily. However, in this work, it is
part of the optimization process. Currently, the literature covering
optimization related to the choice of the window size period is very
limited. Although some studies investigate this in terms of sensitivity
analysis, for example, \textcite{Karathanasopoulos2016} used two different
windows for the testing period: 300 days and 400 days. In other
research, \textcite{krynska2023daily} investigated
the effect of varying in-sample and out-of-sample lengths on their
trading strategy results. They found that extending the training period
(in-sample) while decreasing the testing period (out-of-sample) resulted
in improved performance.

The strategy built in this study uses multiple techniques tailored to
its specific needs, including:

\begin{itemize}
\tightlist
\item
  Walk-forward method optimization
\item
  Technical analysis indicator Exponential Moving Average (EMA) with
  optimized length of MA periods.
\item
  Bootstrap
\end{itemize}

\hypertarget{the-walk-forward-method}{%
\subsection{The Walk-Forward Method}\label{the-walk-forward-method}}

This research employed walk-forward optimization (WFO) to ensure that all reported results came from out-of-sample data - either from WFO testing periods or single-time evaluations on the Unseen Data Period. This
approach limits the probability of overfitting, which occurs when
strategies optimized on in-sample data fail to generalize to new market conditions.

The WFO method is just an adaptation of a popular technique used in data
science and quantitative finance, where the validity and the robustness
of model parameters are typically evaluated by dividing the data into
different sets:

-- Training set (in-sample data): This data is used to train the model
and fine-tune its parameters - Validation set (optional in some cases):
This data controls the training process on data that has not been seen by the model - Testing set (out-of-sample data): This set of data provides the final assessment of the model's capabilities to generate good
performance on truly new data.

The walk-forward optimization technique is a method used to evaluate the
robustness and potential profitability of trading strategies. It
involves dividing the historical data into n equal periods. Each of these n periods is further divided into two or three when a validation set is included, into segments with different lengths:

\begin{itemize}
\item
  In-sample/training period: This segment is used to optimize the
  parameters of the trading strategy. Essentially, the strategy is
  ``trained'' on this data to find the settings that yield the best
  performance. In this study, the training period is used to find a combination of fast/slow lengths of Exponential Moving Average that maximizes Sharpe.
\item
  Out-of-sample/testing period: This segment is used to validate the
  performance of the strategy using the parameters found in the
  in-sample period. This acts as a simulation of applying the strategy
  to the new set of data, helping to avoid overfitting and assess its
  adaptability to different market conditions. In the context of this
  study, the out-of-sample period evaluates a combination of strategy
  parameters - fast/slow of the Exponential Moving Average found in the
  in-sample period.
\end{itemize}

Combining results: Only the results from all out-of-sample periods are
used to calculate overall statistics for the global training period.
This prevents overfitting by avoiding the use of in-sample (training) data in performance evaluation.

The impact of using different training and testing data lengths within the walk-forward optimization process on strategy performance was further investigated and detailed in the following sections.

\hypertarget{exponential-moving-average-crossover}{%
\subsection{Exponential Moving Average
Crossover}\label{exponential-moving-average-crossover}}

It is a momentum strategy using moving averages to generate buy and sell
signals. The simplest moving average crossover strategy uses two
averages of prices. One is shorter and sometimes called fast, while the other one is longer and called slow. Crossover occurs when a short-term moving average crosses through a long-term moving average. When a crossover occurs, practitioners believe the momentum of the price is shifting its
direction.

This study employed a set of Exponential Moving Averages (EMAs) with
periods of 5, 7, 10, 15, 20, 30, 40, 50, 100, 150, and 200. EMAs with
periods longer than 35 were categorized as slow, while those with
shorter periods were classified as fast. The selection of this threshold
was somewhat arbitrary.

Formula \eqref{eq:ema_formula} below presents the recursive form of the Exponential Moving Average:

\begin{equation} 
  EMA_\text{n} = \alpha * P_n + (1-\alpha) * EMA_\text{n-1}
  \label{eq:ema_formula}
\end{equation}

\begin{align*} 
\text{where :} \\
n        &- \text{Number of periods for which EMA should be calculated} \\
P_n      &- \text{Current asset price}  \\
\alpha   &= \frac{2}{n+1} \\
EMA_0   &= P_0
\end{align*}

\hypertarget{position-of-strategy}{%
\subsubsection{Position of strategy}\label{position-of-strategy}}

For a specific combination of fast and slow Exponential Moving Average
(EMA) parameters, the strategy:

\begin{itemize}
\tightlist
\item
  Enters long positions (buy) when the fast EMA crosses above the slow
  EMA. This indicates a potential upward trend, suggesting the price
  might continue to rise, and the strategy aims to capitalize on this
  movement.
\item
  Enters short positions (sell) when the fast EMA crosses below the slow
  EMA. This suggests a potential downward trend, and the strategy aims
  to profit from a price decline.
\end{itemize}

The logic for position calculation could be summarized in the formula
\eqref{eq:position_formula}:

\begin{equation} 
  EMA_\text{Position} = \begin{cases}
    \text{Long},& \text{if } EMA_\text{fast}\geq EMA_\text{slow}\\
    \text{Short},   & \text{otherwise}
\end{cases}
 \label{eq:position_formula}
\end{equation}

\hypertarget{optimization}{%
\subsection{Optimization}\label{optimization}}

This study went beyond optimizing EMA parameters and introduced an additional layer of optimization by fine-tuning the training/testing
period length for the walk-forward, potentially leading to a more robust
strategy.

\hypertarget{specific-periods-of-walk-forward-trainingtesting-considered}{%
\subsubsection{Specific periods of walk-forward training/testing
considered}\label{specific-periods-of-walk-forward-trainingtesting-considered}}

To investigate the presence of potential weekly seasonality, in addition to shorter periods of 1, 2, 3, 5, 7, and 10 days, the study explored
various training/testing WF period combinations using multiples of 7
days. These included periods of 14, 21, and 28 days.

\hypertarget{optimization-approach}{%
\subsubsection{Optimization approach}\label{optimization-approach}}

Each combination of training and testing window sizes underwent a
separate walk-forward optimization process. This process used the entire
global training period to assess the impact of specific window sizes on
the strategy's performance. Each iteration of the walk-forward procedure
with a unique window size combination produced a Sharpe ratio, which was
then stored for further analysis. Importantly, the Sharpe Ratio was
calculated based on the walk-forward returns, specifically derived from
the test phase of each of the walk-forward steps.

\hypertarget{visualization}{%
\subsubsection{Visualization}\label{visualization}}

\begin{figure}[H]
\includegraphics{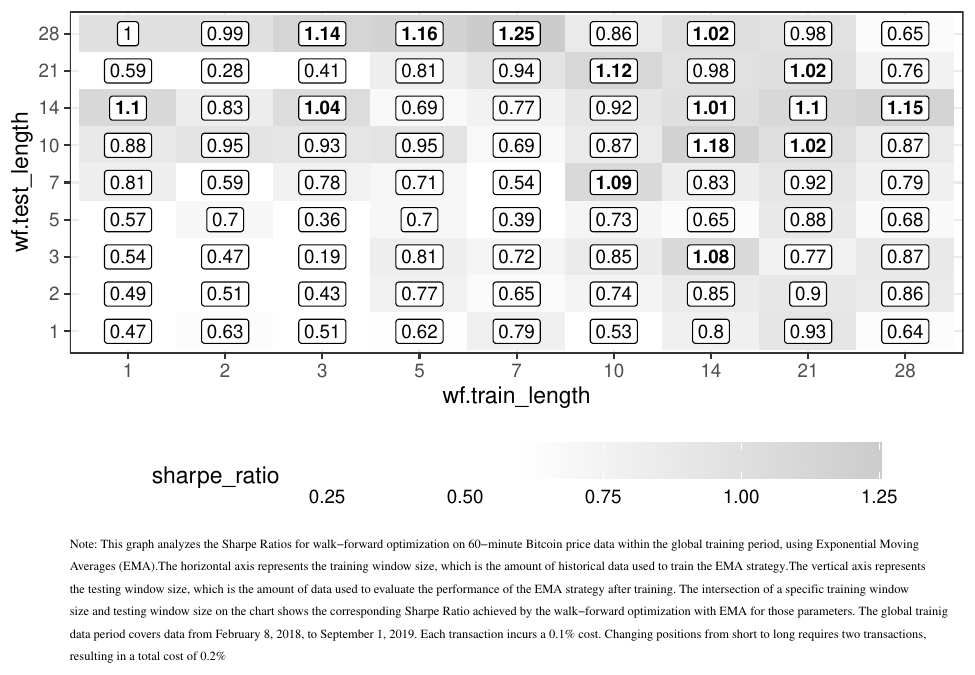} \caption{\label{sharpe_grid}Walk-Forward Sharpe Ratios: Heatmap by Training/Testing Length}\label{fig:sharpe_heatmap }
\end{figure}

The overall results of the above calculation can be presented as a grid
in Figure \ref{sharpe_grid}. The horizontal axis represents the training
window size, and the vertical axis represents the test window size. Each
intersection point displays the Sharpe ratio achieved using that
specific combination for the global training period.

\hypertarget{smoothing-grid}{%
\subsubsection{Smoothing grid}\label{smoothing-grid}}

The grid presented in Figure \ref{sharpe_grid} reveals clusters of
higher and lower Sharpe ratios across different combinations of training
and testing periods. However, directly choosing a single set of
parameters is challenging due to the high variability in results in
neighboring states. To address this variability, a weighted smoothing
method was employed. This method weights the Sharpe ratio of a cell
based on the average Sharpe ratios of its neighboring cells, effectively
giving more weight to areas with consistent positive performance.

\hypertarget{choice-of-smoothing-weights}{%
\subsubsection{Choice of Smoothing
Weights}\label{choice-of-smoothing-weights}}

While the specific choice of smoothing weights can be subjective, in
this case, they were chosen following the method described in
\textcite{gomez2021robust}. Smoothing is achieved
by using the formula \eqref{eq:smoothing} for a weighted average where the
original value receives half the weight, and the remaining weight is
distributed among neighboring values.\\
\begin{equation} 
  Smoothed_\text{value} = \frac{1}{2}*Original_\text{value}  +\frac{1}{2}*\sum_{n=1}^{n}\frac{1}{n}Neighbor_\text{value}
  \label{eq:smoothing}
\end{equation}

\hypertarget{identifying-optimal-parameters-for-out-of-sample-evaluation-based-on-the-smoothed-grid}{%
\subsubsection{Identifying optimal parameters for out-of-sample
evaluation based on the smoothed
grid}\label{identifying-optimal-parameters-for-out-of-sample-evaluation-based-on-the-smoothed-grid}}

\begin{figure}[H]
\includegraphics{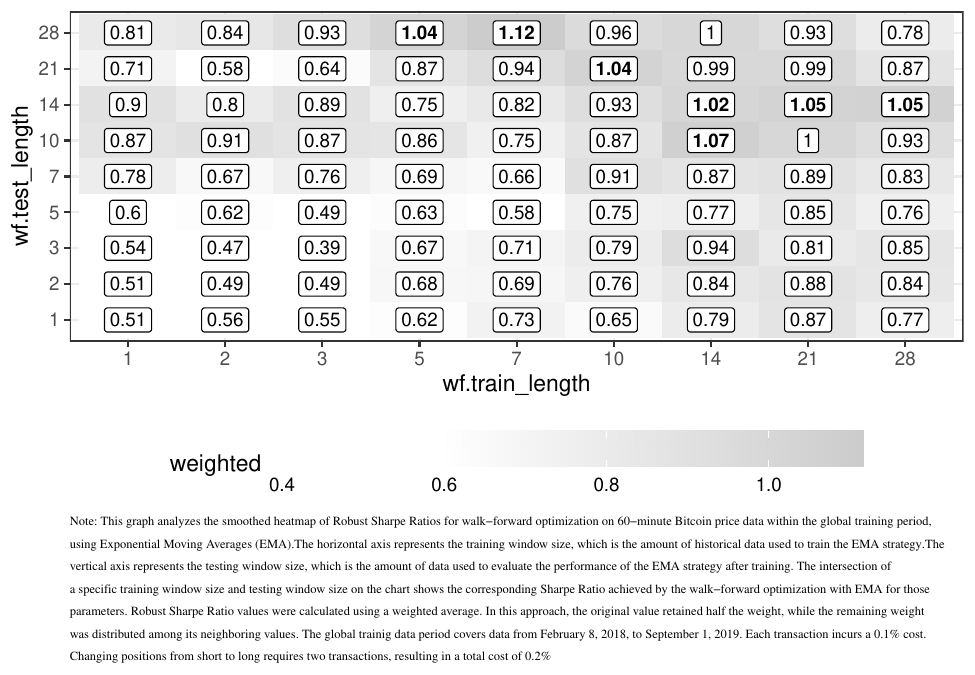} \caption{\label{sharpe_grid_smoothed}Walk-Forward Robust Sharpe Ratios: Heatmap (Smoothed Values) by Training/Testing Length}\label{fig:sharpe_heatmap_smoothed}
\end{figure}

After performing the smoothing procedure described in the previous
paragraph, the two walk-forward parameter combinations with the highest
Sharpe ratios in the smoothed grid (Figure \ref{sharpe_grid_smoothed})
are chosen for further analysis. This selection method is objective and
based on the chosen performance metric (Sharpe ratio). However, the study acknowledges that this approach might not be the most effective,
and alternative methods for selecting optimal walk-forward parameters
could be explored. Alternative methods could include:

\begin{itemize}
\tightlist
\item
  Minimizing the difference between training and testing period results.
  This approach aims to select parameters that lead to similar
  performance in both training and testing phases, suggesting better
  generalizability to unseen data.
\item
  Using bootstrapping to select the walk-forward parameter combination
  with the highest statistical significance. Bootstrapping is a
  resampling technique that can help assess the reliability of the
  parameter selection process.
\end{itemize}

\hypertarget{overall-optimization-summary}{%
\subsubsection{Overall optimization
summary}\label{overall-optimization-summary}}

This optimization process effectively assessed the impact of different
training and testing period combinations on the strategy's performance,
providing valuable insights for selecting a more robust parameter set.
However, it is important to acknowledge potential limitations:

\begin{itemize}
\item
  Data mining: Analyzing numerous training/testing window size
  combinations, combined with selecting the best EMA within the
  walk-forward training period, raises concerns about data mining. This
  practice can lead to overfitting, where the chosen parameters perform
  well on the specific data used but may not generalize well to unseen
  data.
\item
  Training/unseen data split: The chosen split between the global
  training and unseen data periods can also influence parameter
  selection. Different splits could potentially lead to different
  optimal parameters.
\item
  Parameter decay: The optimization process does not consider the
  possibility that the chosen parameters might become less effective
  over time. The strategy's performance on unseen data could potentially
  deteriorate if the parameters found in the global training period are
  no longer optimal.
\end{itemize}

These limitations highlight the need for further investigation into
robust parameter selection techniques and the potential for performance
decay over time.

\hypertarget{statistical-significance}{%
\subsection{Statistical significance}\label{statistical-significance}}

This study employed bootstrapping to evaluate statistical significance.
However, as noted by \textcite{Ruiz2002}, the classical bootstrap
introduced by \textcite{Efron1977} is not suitable for time
series data due to potential heteroscedasticity and autocorrelation.
Therefore, the block bootstrap method developed by \textcite{Politis1994} was customized to specifically address these concerns and better fit the nature of time series data, and was the method used in this paper.

\hypertarget{bootstrap-evaluation-of-subsampled-ema-combinations-for-walk-forward-optimization}{%
\subsubsection{Bootstrap Evaluation of Subsampled EMA Combinations for
Walk-Forward
Optimization}\label{bootstrap-evaluation-of-subsampled-ema-combinations-for-walk-forward-optimization}}

Within each step of the walk-forward training period, the study
evaluated various combinations of exponential moving averages (EMAs).
The combination that generated the highest Sharpe Ratio during the
training period was then chosen for further assessment on the subsequent
walk-forward testing period. In contrast, the bootstrapping method
employed a different approach. Instead of selecting the best EMA
combination based on the training data, it utilized random selection.
This effectively compared the performance of the ``best'' parameter
choice, identified during the walk-forward optimization, against a
baseline established by randomly chosen combinations. Due to the
computationally intensive nature of this method (requiring 1,000
iterations), it was only applied to the top two train/testing period
combinations identified through their Sharpe Ratio during the
walk-forward process on the global testing period.

\hypertarget{bootstrap-evaluation-of-shuffled-transaction-blocks}{%
\subsubsection{Bootstrap Evaluation of Shuffled Transaction
blocks}\label{bootstrap-evaluation-of-shuffled-transaction-blocks}}

The transactional positions bootstrap was another method of bootstrap that was
adapted for this study. During the walk-forward testing
period, ``position blocks'' were generated. These blocks represented
consecutive periods where the strategy held long or short positions
(e.g., 3 periods long, followed by 2 short, then 5 long). In each
bootstrap iteration, the order of these blocks was shuffled. For
example, the 2-period short block might be swapped to come before the
8-period long block (5 + 3), and then original asset returns are matched
to each block to construct the equity line and calculate profitability
for a single iteration. Due to the computationally intensive nature of
this method (requiring 1,000 iterations), it was only applied to the top
two train/testing period combinations identified through their Sharpe
Ratio during the walk-forward process on the global testing period.

\hypertarget{null-hyphotestis}{%
\subsubsection{Null hyphotestis}\label{null-hyphotestis}}

Both block and moving average (MA) bootstrapping methods assumed a null
hypothesis that the research strategy does not provide any statistically significant advantage over randomly generated trading signals. In
simpler terms, the hypothesis assumed the strategy's performance is
purely due to chance.

To reject this null hypothesis and conclude that the strategy has a
statistically significant edge, it needs to outperform a predetermined
proportion of random strategies based on the chosen confidence level.

Confidence level represents the probability that the observed effect
(strategy's edge) didn't occur by chance. A higher confidence level
indicates a stronger likelihood that the effect is real.

For example, with a 5\% confidence level, the strategy needs to be
demonstrably better than 95\% of randomly generated strategies to be
considered statistically significant. This implies that the strategy's
performance should consistently outperform 95\% of randomly generated
signals, suggesting it has a genuine advantage over a random approach.

\hypertarget{cost-sensitivity}{%
\subsection{Cost sensitivity}\label{cost-sensitivity}}

While the influence of transaction costs on the strategy outcomes has received limited academic attention, this study delved into this
subject. One of the previous studies by \textcite{Svogun2022} demonstrated the significant impact of transaction costs on trading strategies using intraday data
(1-minute), while having a minimal effect on strategies using 1-day
data.

Transaction costs are a critical issue, especially for moving average
strategies. These strategies hold positions continuously, leading to
frequent position changes and higher sensitivity to transaction fees. A
single long-to-short turnaround with the default cost of 0.1\% incurs a
total cost of 0.2\% (entry + exit). This cost scales proportionally with
the transaction fee level, reaching 1\% at the highest cost of 0.5\%
analyzed in this research

This study used a 0.1\% cost assumption for all calculations. This fee
level is not tied to a specific exchange rate but reflects a commonly
observed level offered by the largest cryptocurrency exchanges.

While some exchanges may offer lower fees under specific circumstances, the 0.1\% assumption takes into account not only transaction fees but
also potential slippage and changing bid-ask spread. This broader
consideration aims to provide a realistic estimate of trading costs in
real-world scenarios.

To assess the impact of transaction costs on the strategy's performance, this study analyzed results under seven different cost levels: 0.05\%,
0.07\%, 0.10\%, 0.20\%, 0.30\%, 0.40\%, 0.50\%

This study investigated the impact of transaction costs on the
performance of the Exponential Moving Average Crossover strategy. The
strategy generated signals to enter long or short positions, and each
position change incurred a transaction cost. Notably, switching from a
long position to a short position requires closing the long position
before opening the short, resulting in double value of the transaction
cost for such a turnaround.

To assess cost sensitivity, the analysis applied varying transaction cost levels throughout the global training period to the original
positions list generated by the strategy. While this approach might not
perfectly replicate real-world scenarios where higher costs incentify strategies with fewer trades, it provides a valuable starting point for understanding how transaction costs affect high-frequency strategy performance with all other factors held constant.

\hypertarget{algorithms-descriptions}{%
\subsection{Algorithms descriptions}\label{algorithms-descriptions}}

This section delves into the details of the algorithm employed in the
study. The algorithm itself is presented in pseudocode format

\hypertarget{wf-parameters-optimization-algorithm-with-evaluation-of-results-on-the-unseen-period}{%
\subsubsection{WF parameters optimization algorithm with evaluation of
results on the unseen
period}\label{wf-parameters-optimization-algorithm-with-evaluation-of-results-on-the-unseen-period}}

Algorithm \ref{alg:wf_param_opt} outlines the steps involved in
optimizing parameters for the walk-forward optimization process. To
enhance clarity, this algorithm has been segmented into two distinct
parts. The first part, presented here, focuses on the optimization of
training and testing window sizes within the walk-forward framework. The
second part, detailed in section
\protect\hyperlink{the-walk-forward-procedure-algorithm}{The Walk-Forward procedure algorithm}, describes the actual walk-forward
procedure itself.

\begin{enumerate}
\def\labelenumi{\arabic{enumi}.}
\tightlist
\item
  Looping Through Parameter Combinations: The algorithm iterates through
  all possible combinations of training and testing window sizes, which are the key parameters in the optimization process. This process is
  described in Algorithm \ref{alg:wf_param_opt}.
\item
  Walk-Forward Execution: The walk-forward procedure is executed for
  each parameter combination. This step constitutes the core trading
  logic, involving the calculation of the EMA at each walk-forward step.
  A detailed description of the walk-forward algorithm can be found in Algorithm \ref{alg:wf_procedure}.
\item
  Sharpe Calculation: The performance of each walk-forward evaluation is
  measured using the Sharpe Ratio, a metric that considers both returns
  and risk.
\item
  Sharpe Ratio Grid Construction: Once all combinations are tested, the
  algorithm constructs a ``grid'' using the obtained Sharpe Ratios. Each
  column in this grid represents the training window size used, while
  each row represents the testing window size.
\item
  Smoothing the Sharpe Ratio Grid: This step involves refining the
  Sharpe Ratio grid by applying a smoothing technique. This might
  involve incorporating both the original Sharpe Ratio value and the
  values of its neighboring cells in the grid into the calculation.
\item
  Selecting Optimal Combinations: Finally, the two combinations of
  training and testing window size that lead to the highest Sharpe
  Ratios in the smoothed grid are chosen.
\item
  Applying Strategy to Unseen Data Period: These two chosen parameter
  combinations are then used to evaluate the walk-forward strategy on
  the Unseen Data Period for three specific cryptocurrencies: Bitcoin
  (BTC), Binance Coin (BNB) and Ethereum (ETH).
\end{enumerate}

\singlespacing

\begin{algorithm}
\caption{WF Parameters opitmization  }\label{alg:wf_param_opt}
\begin{algorithmic}[1]

\Procedure{WFParamtersOptimization}{$DataSeries$}
    \State Divide $DataSeries$ into $globalTrain$, $unseenTest$
    \State Define $trainCoin$ used in global train optimization 
    \State Define $testCoins$ used in unseenTest
    \State Define $WFPairs(train,test)$ which consists of a combination of Walk-Forward train/test length period
    \State Define matrix $WFGlobalTrainSharpe$ which will store Sharpe for each combination of train/test
    \For{each WFPair($train$, $test$) in $WFPairs$}
        \State Call $WFCalculateSharpe(globalTrain for $trainCoin$,WFPair(train,test))$ procedure and store results in $WFGlobalTrainSharpe$ 
    \EndFor
    \State From $WFGlobalTrainSharpe$ get 2 $MaxWFPairs(train,test)$ with highest Sharpe
    \For{each WFPair($train$, $test$) in $MaxWFPairs$}
        \State Get $unseenTest$  for $testCoins$
        \State Call $WFCalculateSharpe($unseenTest$,WFPair(train,test))$ procedure and store results in $UnseenResults$ 
    \EndFor
    \State Report $UnseenResults$
\EndProcedure

\end{algorithmic}
\end{algorithm}
\onehalfspacing

\hypertarget{the-walk-forward-procedure-algorithm}{%
\subsubsection{The Walk-Forward procedure
algorithm}\label{the-walk-forward-procedure-algorithm}}

The walk-forward procedure is executed for each combination of
training/testing window sizes during the walk-forward parameter
optimization process, which is detailed in section \ref{alg:wf_param_opt}.

\begin{enumerate}
\def\labelenumi{\arabic{enumi}.}
\tightlist
\item
  Data Segmentation: Involves segmenting the historical data series into
  walk-forward steps. Each walk-forward step has a length equal to the
  combined size of the training window and the testing window. The   specific combination of training window size and testing window size
  is provided as a parameter when defining the walk-forward optimization
  process.
\item
  Parameter Definition: Define combinations of fast and slow EMA window sizes for the Exponential Moving Average (EMA) strategy. These
  parameters will be optimized for each walk-forward training period.
\item
  Iterating through walk-forward steps: The procedure iterates through
  each data walk-forward step containing both a training period and a
  testing period.
\item
  Training Data Optimization: Within each segment, calculate returns
  using the EMA strategy for various combinations of fast and slow EMA
  window sizes applied to the training data.
\item
  Performance Evaluation: For each combination of EMA periods, calculate the Sharpe Ratio based on the strategy's returns on the training data.
\item
  Optimal Combination Selection: After evaluating all EMA parameter combinations, choose the one that leads to the highest Sharpe Ratio as
  the optimal combination for that specific walk-forward training
  period.
\item
  Testing with Optimal Parameters: Using the chosen EMA parameters (fast
  and slow periods), calculate the strategy's returns on the testing
  data within the current walk-forward step.
\item
  Test Result Accumulation: Store the returns achieved in the testing
  period for further analysis.
\item
  Segment-wise Repetition: Repeat steps 3-8 for all steps in the
  walk-forward process.
\item
  Overall Performance Calculation: Finally, calculate the overall Sharpe
  Ratio for the entire walk-forward process, considering the combined returns from all testing periods across all walk-forward steps.
\end{enumerate}

\singlespacing

\begin{algorithm}
\caption{walk-forward (WF) Procedure }\label{alg:wf_procedure}
\begin{algorithmic}[1]

\Procedure{WFCalculateSharpe}{$DataSeries$,$train$,$test$}
    \State define $MAPairs(fast,slow)$ 
    \State Divide $DataSeries$ into $WFSegments$ with length $train$ + $test$
    \State Initialize  list $WFReturns$  to keep overall returns of the whole WF
    \For{each segment($WFTrain$, $WFTest$) in $WFSegments$}
        \State Define empty $SharpeList$ for keeping the Sharpe ratio for each MA combination on the testing period 
        \For{each $MaPair(fast,slow)$  in $MaPairs$}
          \State calculate $EmaReturns$ using $MaPair(fast,slow)$ on $WFTrain$
          \State calculate $Sharpe$ using $EmaReturns$
          \State store $Sharpe$ with $MaPair(fast,slow)$ in the $SharpeList$
        \EndFor
        \State End For
        \State Get $BestMaPair(fast,slow)$ from $SharpeList$ with highest $Sharpe$
        \State calculate $EMATestReturn$ using $BestMaPair(fast,slow)$ on $WFTest$
        \State calculate $SharpeTest$ using $EMATestReturns$
        \State append $EMATestReturns$ to $WFReturns$
    \EndFor
    \State End For
    \State calculate $WFTotalSharpe$ using $WFReturns$
    \State return $WFTotalSharpe$ 
\EndProcedure

\end{algorithmic}
\end{algorithm}
\onehalfspacing

\newpage

\hypertarget{performance-metrics}{%
\subsubsection{Performance metrics}\label{performance-metrics}}

The following metrics are used across the study to measure profitability
and risk.

Appropriate performance statistics presented in Table~\ref{tab:performance-metrics}
were selected based on \textcite{Rys2019206229}, which extensively describes the process
of creating automated trading systems and the evaluation of strategies' performance.

\begin{table}[!ht]
\centering
\caption{Performance statistics definitions.}
\label{tab:performance-metrics}
\resizebox{\columnwidth}{!}{%
\begin{tabular}{llp{6cm}}
\toprule
Metric & Formula & Additional information \\
\midrule
Log return
& $R_t = ln(\dfrac{p_t }{p_{t-1}})$
& $p_t$ the asset price at time t \\[1em]

Annualized Mean Return

& $(\frac{1}{n}\sum_{t=1}^{N}{R_t} )* n_{year}$
& $n_{year}$ Number of log returns in the year \newline
$N$ number of log returns in the strategy \\[2em]

Annualized Volatility

& $\sqrt{\frac{1}{N-1} \sum_{t=1}^{N}(R_t-\bar{R})} * \sqrt{n_{year}}$
& $\overline{R}$ average log return \\[2em]

Sharpe ratio
& $\frac{\text{Annualized Mean Return} - r_\text{risk free}}{\text{Annualized Volatility}}$
& $r_\text{risk free}$ Risk Free Rate in this study is assumed to be 0 
\\[2em]

Maximum Drawdown
& $ \max_{\tau\in[0,T)}  \max_{t\in[\tau,T]}({p_\tau-p_t,0})$
& Worst cumulative loss from a local maximum \newline 
$p_t $ logarithm of asset price $ln(P_t)$ at the time t and   $t\in[0,T]$ \newline
$P_t$ \text{Asset price at time t }\\[2em]

Information Ratio**
& $\frac{sign(\text{Annualized Mean Return}) * \text{Annualized Mean Return}^{2}}{\text{Annualized Volatility} * \text{Maximum Drawdown}}$
& \\[2em]

Sortino Ratio
& $\frac{\bar{R} - r_\text{benchmark}}{\sqrt{\sigma_\text{Negative log Returns}}} * \sqrt{n_{year}}$
& $r_{benchmark}$    Benchmark return in this study assumed to be 0 \newline 
$\sigma_\text{Negative log Returns}$  Standard deviation of downside log returns, lower than the assumed level of $r_{benchmark}$  \\
\\[2em]
\bottomrule
\end{tabular}%
}
\end{table}

\hypertarget{results---global-training-data-period}{%
\section{Results - Global Training Data
Period}\label{results---global-training-data-period}}

\hypertarget{choice-of-data-frequency-on-results}{%
\subsection{Choice of data frequency on
results}\label{choice-of-data-frequency-on-results}}

This study employed an exhaustive search to identify the optimal
walk-forward (WF) length for each timeframe tested. Essentially, for
every timeframe (1, 5, 10, 15, 30, and 60 minutes), the walk-forward was
optimized across all predefined training and testing period length
combinations. For each interval, the walk-forward performance was
estimated using the same combination of training and testing lengths for
all window sizes (1, 2, 3, 5, 7, 10, 14, 21, 28). This resulted in 81
strategies per timeframe. Each combination yielded a Sharpe Ratio,
reflecting the strategy's risk-adjusted performance. Table
\ref{trad_stats_by_tf} summarizes the descriptive statistics of these
Sharpe ratios for each timeframe. This provides a comprehensive overview
of the strategy's performance across different sampling frequencies.

Key findings:

\begin{itemize}
\tightlist
\item
  Higher frequency timeframe (1-15) might be unprofitable due to 0.1\% transaction costs, which were assumed
\item
  Among the tested timeframes, 15- and 30-minute intervals contain some
  WF combinations with positive returns.
\item
  Notably, the 60-minute timeframe also shows a positive mean Sharpe
  ratio
\item
  No negative Sharpe at 60-minute frequency, even if the strategy is chosen randomly in 60 60-minute timeframe, it still exhibits a positive Sharpe Ratio
\end{itemize}

\begin{table}[H]

\caption{\label{tab:sharpe_per_tf}\label{trad_stats_by_tf} The Descriptive Statistics of Sharpe Ratio for 81 walk-forward combinations per timeframe - Global Train Data Period}
\centering
\fontsize{10}{12}\selectfont
\begin{threeparttable}
\begin{tabular}{rrrrrrrr}
\toprule
\makecell[r]{Frequency\\Minutes} & \makecell[r]{Mean\\Sharpe} & \makecell[r]{Max\\Sharpe} & \makecell[r]{Min\\Sharpe} & \makecell[r]{STD\\Sharpe} & \makecell[r]{Q25\%\\Sharpe} & \makecell[r]{Q50\%\\Sharpe} & \makecell[r]{Q75\%\\Sharpe}\\
\midrule
\cellcolor{gray!6}{1} & \cellcolor{gray!6}{-12.7144} & \cellcolor{gray!6}{-11.4142} & \cellcolor{gray!6}{-16.6509} & \cellcolor{gray!6}{1.0943} & \cellcolor{gray!6}{-12.7549} & \cellcolor{gray!6}{-12.5585} & \cellcolor{gray!6}{-12.1700}\\
5 & -2.8416 & -1.6011 & -5.0528 & 0.8268 & -3.0947 & -2.6447 & -2.3863\\
\cellcolor{gray!6}{10} & \cellcolor{gray!6}{-1.4896} & \cellcolor{gray!6}{-0.4917} & \cellcolor{gray!6}{-2.9345} & \cellcolor{gray!6}{0.6626} & \cellcolor{gray!6}{-1.9249} & \cellcolor{gray!6}{-1.4718} & \cellcolor{gray!6}{-1.0263}\\
15 & -0.9783 & 0.1951 & -2.7203 & 0.7177 & -1.1952 & -0.7957 & -0.5273\\
\cellcolor{gray!6}{30} & \cellcolor{gray!6}{-0.4954} & \cellcolor{gray!6}{0.7372} & \cellcolor{gray!6}{-2.1529} & \cellcolor{gray!6}{0.6440} & \cellcolor{gray!6}{-0.9490} & \cellcolor{gray!6}{-0.4528} & \cellcolor{gray!6}{-0.1158}\\
\addlinespace
60 & 0.7908 & 1.2524 & 0.1863 & 0.2256 & 0.6479 & 0.8124 & 0.9540\\
\bottomrule
\end{tabular}
\begin{tablenotes}[para]
\small
\item \textit{\tiny{ Note: }} 
\item \tiny{ The descriptive statistics of the Sharpe Ratio achieved by Exponential Moving Averages (EMA) strategies with walk-forward optimization executed on different time frequencies within the global training data, using varying training and testing window sizes. For each time frequency, the walk-forward optimization was performed multiple times with the same window combinations. The global training data period covers data from February 8, 2018, to September 1, 2019. Each transaction incurs a 0.1\% cost. Changing positions from short to long requires two transactions, resulting in a total cost of 0.2\% }
\end{tablenotes}
\end{threeparttable}
\end{table}

\hypertarget{the-walk-forward-optimization-results}{%
\subsection{The walk-forward optimization
results}\label{the-walk-forward-optimization-results}}

\hypertarget{walk-forward-sharpe-ratios-heatmap-by-trainingtesting-length-60-minute}{%
\subsubsection{Walk-Forward Sharpe Ratios: Heatmap by Training/Testing
Length
(60-Minute)}\label{walk-forward-sharpe-ratios-heatmap-by-trainingtesting-length-60-minute}}

Figure \ref{sharpe_grid} presents a heatmap visualizing the Sharpe
ratios achieved through a walk-forward optimization process applied to
60-minute Bitcoin price data within the global training dataset. The
columns represent the lengths of the training periods, while the rows
represent the lengths of the testing periods. Each cell in the heatmap
displays the Sharpe ratio achieved using a specific combination of
training and testing period lengths within the global training data.

The heatmap could be divided into four quarters:

First Quadrant (Bottom-Left):

\begin{itemize}
\tightlist
\item
  This quadrant corresponds to short walk-forward testing lengths
  (around 1 to 7 days) and short walk-forward training lengths (around 1
  to 7 days).
\end{itemize}

Key insights:

\begin{itemize}
\tightlist
\item
  Generally low Sharpe ratios.
\item
  Best training period: 5 days.
\item
  Best testing period: 7 days.
\item
  Surprisingly, combinations \textless= 3 days yield the lowest Sharpe
  ratios, suggesting frequent retraining might not always be beneficial.
\item
  For 1-day training, testing periods of 5-7 days show the best results
\end{itemize}

Second Quadrant (Top-Left):

\begin{itemize}
\tightlist
\item
  Walk-forward training lengths (around 1 to 7 days) and longer
  walk-forward testing lengths (around 8 to 28 days).
\end{itemize}

Key insights

\begin{itemize}
\tightlist
\item
  The highest Sharpe ratios across all training lengths were observed for testing
  periods of 28 days.
\item
  Best training periods: 5-7 days.
\item
  Best performance: 7-day training and 28-day testing.
\end{itemize}

Third Quadrant (Bottom-Right):

\begin{itemize}
\tightlist
\item
  Walk-forward training lengths (around 10 to 28 days) and longer
  walk-forward testing lengths (around 1 to 7 days).
\end{itemize}

Key insights:

\begin{itemize}
\tightlist
\item
  Only 2 combinations with training periods of 10 and 14 days have
  Sharpe ratios exceeding 1.
\item
  1-day testing period consistently yields the lowest Sharpe ratios.
\end{itemize}

Fourth Quadrant (Top-Right):

\begin{itemize}
\tightlist
\item
  Walk-forward training lengths (around 10 to 28 days) and longer
  walk-forward testing lengths (around 10 to 28 days)
\end{itemize}

Key insights:

\begin{itemize}
\tightlist
\item
  8 combinations achieve Sharpe ratios above 1.
\item
  Best performance: 14-day training and 10-day testing.
\end{itemize}

Overall: Quadrant 4 (long training, long testing) surprisingly exhibits
the highest Sharpe ratios. This finding seems counterintuitive because
less frequent retraining might prevent the strategy from adapting
quickly to changing market conditions. Further investigation is
necessary to understand this outcome.

\hypertarget{walk-forward-robust-sharpe-ratios-heatmap-smoothed-values-by-trainingtesting-length-60-minute}{%
\paragraph{\texorpdfstring{Walk-Forward Robust Sharpe Ratios: Heatmap
(Smoothed Values) by Training/Testing Length (60-Minute)
\newline}{Walk-Forward Robust Sharpe Ratios: Heatmap (Smoothed Values) by Training/Testing Length (60-Minute) }}\label{walk-forward-robust-sharpe-ratios-heatmap-smoothed-values-by-trainingtesting-length-60-minute}}

Figure \ref{sharpe_grid_smoothed} presents a heatmap with a Robust Sharpe
Ratio derived from the data in Figure \ref{sharpe_grid}. The smoothing
formula proposed by \textcite{gomez2021robust} is described
in section \protect\hyperlink{choice-of-smoothing-weights}{Choice of
Smoothing Weights}. This technique takes into account not only each
value's original position but also the values of its neighboring cells,
resulting in a clearer visualization of overall trends. Following the work of \textcite{gomez2021robust}, the formula for the calculation of the Robust Sharpe Ratio for a given cell in the heatmap
is presented here \eqref{eq:robust_sharpe}:

\begin{equation} 
  \text{Robust Sharpe Ratio} = \frac{1}{2}*\text{Sharpe Ratio}_\text{ Original}  +\frac{1}{2}*\sum_{n=1}^{n}\frac{1}{n}\text{Sharpe Rato }_\text{Neighbour}
  \label{eq:robust_sharpe}
\end{equation}

\begin{align*} 
\text{where :} \\
n                                         &- \text{the index of the neighbor.} \\
\text{Sharpe Ratio}_\text{ Original}      &- \text{Sharpe Ratio for given combination of training/testing period} \\
\sum_{n=1}^{n}\frac{1}{n}\text{Sharpe Rato }_\text{Neighbour}     &- \text{The sum of all neighbors} 
\end{align*}

Key insights:

\begin{itemize}
\tightlist
\item
  The highest Sharpe Ratios are observed in the area with longer
  training periods (14-21-28 days), specifically for 14-days and 10-days
  testing periods
\item
  Based on this Robust Sharpe Ratio heatmap, two combinations were
  selected for further testing on Unseen Data Period:

  \begin{itemize}
  \tightlist
  \item
    training 7 days testing 28 days
  \item
    training 14 days testing 10 days
  \end{itemize}
\item
  It's important to note that this selection was somewhat subjective, as
  it assumed the best training period performance translates to the best
  performance on the Unseen Data Period. This is not guaranteed.
\item
  Alternative selection criteria could be explored, such as choosing
  combinations with the most stable results across different training
  and testing lengths.
\end{itemize}

\hypertarget{selected-combinations}{%
\paragraph{Selected Combinations}\label{selected-combinations}}

\hfill\break
This study selected parameters for the Unseen Data Period evaluation
based on their performance in the global training period. Specifically, the two parameter sets that achieved the highest smoothed Sharpe Ratios
(presented in Figure \ref{sharpe_grid_smoothed}) were chosen.

However, it is important to acknowledge potential alternative choices.
An astute reader might identify combinations like 21/14 on the smoothed
Sharpe grid, which appears visually more promising due to its location
within the central ``greyish area'' of the heatmap, indicating a higher
smoothed Sharpe ratio compared to the chosen 14/10.

Furthermore, the choice of 7/28 days raises another potential concern.
This selection resides on the edge of the search window, with estimates
based on only 5 neighboring cells instead of the usual 8. This
introduces a higher degree of uncertainty compared to centrally located
choices.

Despite these considerations, the study emphasizes that the parameter
selection process aimed for simplicity, prioritizing the two
best-performing parameter sets based on the initial smoothing pass.

The following combinations achieved the highest Sharpe Ratios in the
Robust Sharpe Ratio Heatmap (presented in Figure
\ref{sharpe_grid_smoothed}) and were chosen for further evaluation:

\begin{itemize}
\tightlist
\item
  Training 7 days, Testing 28 days\\
\item
  Training 14 days, Testing 10 days
\end{itemize}

\hypertarget{performance-metrics---global-training-data-period---walk-forward-top-combinations}{%
\subsection{Performance Metrics - Global Training Data Period -
Walk-Forward Top
Combinations}\label{performance-metrics---global-training-data-period---walk-forward-top-combinations}}

Table \ref{tbl_trade_stats_train} summarizes performance metrics for
trading strategies identified through a walk-forward optimization
process, where the length of the training and testing periods of a
walk-forward was optimized. These strategies, which use Exponential Moving
Averages (EMA) achieved the highest Sharpe Ratio for Bitcoin within the
global training period. The process of selecting of optimal walk-forward
parameters was described in the
\protect\hyperlink{selected-combinations}{Selected Combinations}
section.

Key findings:

\begin{itemize}
\tightlist
\item
  Annualized Mean Return: Both strategies achieved similar returns, with
  a difference of only 5\% compared to the overall 94\% return. This
  difference was considered negligible.
\item
  Annualized Volatility: Both strategies exhibited identical volatility.
\item
  Sharpe Ratio: The 7/28 combination (7 days training, 28 days testing)
  had a higher Sharpe ratio due to its slightly higher annualized
  return.
\item
  Maximum Drawdown: The 7/28 combination significantly outperformed the
  other with a much smaller maximum drawdown, indicating better downside
  risk protection.
\item
  Information Ratio**: The 7/28 combination (training 7 days, testing 28
  days) achieved a higher Information Ratio compared to alternatives,
  driven by its superior performance in both return and drawdown.
\item
  Sortino Ratio: While marginally higher for the 7/28 strategy, this was
  likely related to its better performance in terms of drawdown.
\end{itemize}

\begin{table}[H]

\caption{\label{tab:trading_stats_global_train}\label{tbl_trade_stats_train}Trading Metrics: Top Sharpe Ratio Strategies (Walk-Forward, Global Training)}
\centering
\fontsize{10}{12}\selectfont
\begin{threeparttable}
\begin{tabular}{rrrrrrr}
\toprule
Description & \makecell[r]{Annualized\\Mean\\Return} & \makecell[r]{Annualized\\Volatility} & \makecell[r]{Sharpe\\Ratio} & \makecell[r]{Information\\Ratio**} & \makecell[r]{Max\\Drawdown} & \makecell[r]{Sortino\\Ratio}\\
\midrule
\cellcolor{gray!6}{BTC Train 7 Test 28} & \cellcolor{gray!6}{0.9483} & \cellcolor{gray!6}{0.7572} & \cellcolor{gray!6}{1.252} & \cellcolor{gray!6}{4.622} & \cellcolor{gray!6}{0.3516} & \cellcolor{gray!6}{1.799}\\
BTC Train 14 Test 10 & 0.8918 & 0.7571 & 1.178 & 2.902 & 0.4513 & 1.678\\
\bottomrule
\end{tabular}
\begin{tablenotes}[para]
\small
\item \textit{\tiny{ Note: }} 
\item \tiny{ This table presents trading metrics for strategies identified through a walk-forward optimization process, where the length of the training and testing periods of a walk-forward was optimized. These strategies use Exponential Moving Averages (EMA), 
achieved the highest Sharpe Ratio for Bitcoin within the global training period. The global training data period covers data from February 8, 2018, to September 1, 2019. Each row showcases the performance metrics when using varying training and testing window sizes for a walk-forward optimization. Each transaction incurs a 0.1\% cost. Changing positions from short to long requires two transactions, resulting in a total cost of 0.2\% }
\end{tablenotes}
\end{threeparttable}
\end{table}

\hypertarget{equity-curves---global-training-data-period---walk-forward-top-combinations}{%
\subsection{Equity Curves - Global Training Data Period - Walk-Forward
Top
Combinations}\label{equity-curves---global-training-data-period---walk-forward-top-combinations}}

\begin{figure}[H]
\includegraphics{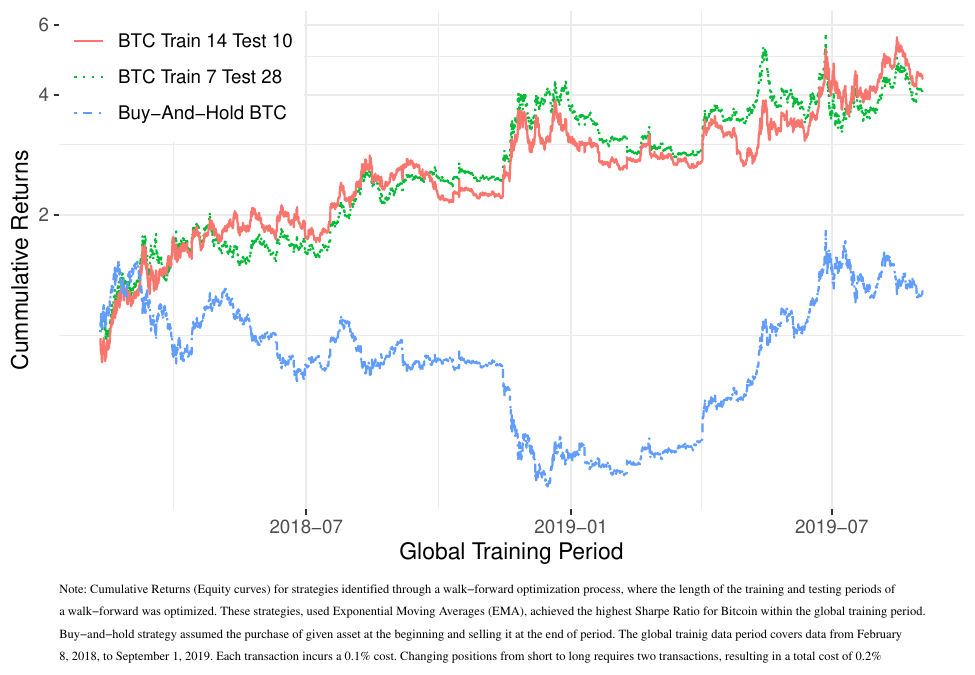} \caption{\label{equity_curve_train}Equity Curves: Top Sharpe Ratio Strategies (Walk-Forward, Global Training Data Period)}\label{fig:unnamed-chunk-1}
\end{figure}

Figure \ref{equity_curve_train} compares the cumulative returns (equity
curves) of a Buy-and-Hold strategy with trading strategies identified
through a walk-forward optimization process, where the length of the training and testing periods of a walk-forward was optimized. These
strategies, using Exponential Moving Averages (EMA), achieved the highest
Sharpe Ratio for Bitcoin within the global training period. Buy-and-Hold strategy assumes the purchase of a given asset at the beginning and
selling it at the end of the period. The process of selecting of optimal walk-forward parameters was described in the \protect\hyperlink{selected-combinations}{Selected Combinations}
section.

Key findings:

\begin{itemize}
\tightlist
\item
  Both EMA strategies exhibited similar overall performance,
  outperforming the Buy-and-Hold strategy.
\item
  The EMA strategies tended to be more profitable when Bitcoin prices
  declined, suggesting they capitalized on short-term trends.
\item
  Conversely, when Bitcoin prices recovered (e.g., after the Bitcoin
  decline in early 2019), the EMA strategies underperformed the
  Buy-and-Hold.
\item
  Most of the gains for the EMA strategies occurred in the first half of
  the period, while the second half saw them lagging behind the
  recovering Buy-and-Hold.
\end{itemize}

\hypertarget{statistical-significance---bootstraps}{%
\subsection{Statistical significance -
Bootstraps}\label{statistical-significance---bootstraps}}

To assess the effectiveness of the chosen parameter combinations for a
walk-forward optimization and evaluate their statistical significance, two bootstrap procedure tests were performed. Details of the procedures
are described in the
\protect\hyperlink{statistical-significance}{Statistical significance}
section.

\hypertarget{results---bootstrap-evaluation-of-subsampled-ema-combinations-for-walk-forward-optimization}{%
\subsubsection{Results - Bootstrap Evaluation of Subsampled EMA
Combinations for Walk-Forward
Optimization}\label{results---bootstrap-evaluation-of-subsampled-ema-combinations-for-walk-forward-optimization}}

Both strategy combinations selected in the
\protect\hyperlink{selected-combinations}{Selected Combinations} section
were evaluated for statistical significance using the method detailed in
the
\protect\hyperlink{bootstrap-evaluation-of-subsampled-ema-combinations-for-walk-forward-optimization}{Bootstrap
Evaluation of Subsampled EMA Combinations for Walk-Forward Optimization}
section. This method differs from the standard approach by using random
EMA parameters drawn from a predefined set described in the section
\protect\hyperlink{exponential-moving-average-crossover}{Exponential
Moving Average Crossover}, whereas the standard approach selects the
best EMA parameters identified during the training period for each
walk-forward step. This allows assessing the performance of the
selection method compared to random selection. A bootstrap procedure was
performed during the global training period.

\begin{table}[H]

\caption{\label{tab:boot_ma_table }Results - Bootstrap Evaluation of Subsampled EMA Combinations for Walk-Forward Optimization}
\centering
\fontsize{10}{12}\selectfont
\begin{threeparttable}
\begin{tabular}{rrrrr}
\toprule
Description & \makecell[r]{Original\\Sharpe Ratio} & \makecell[r]{Total No. of\\bootstrap iterations} & \makecell[r]{No. Bootstrap Iterations\\with Higher Sharpe Ratio} & Significance\%\\
\midrule
\cellcolor{gray!6}{BTC Train 7 Test 28} & \cellcolor{gray!6}{1.252} & \cellcolor{gray!6}{1000} & \cellcolor{gray!6}{80} & \cellcolor{gray!6}{8.0}\\
BTC Train 14 Test 10 & 1.178 & 1000 & 137 & 13.7\\
\bottomrule
\end{tabular}
\begin{tablenotes}[para]
\item \textit{\tiny{ Note: }} 
\item \tiny{ Original Sharpe Ratio - Sharpe Ratios of the Pre-Selected Best Strategies from the Global Training Period,  Total No. of bootstrap iterations - Number of times the bootstrap procedure was repeated, No. Bootstrap Iterations with Higher Sharpe Ratio - Number of bootstrap iterations that achieved a Sharpe Ratio exceeding the original strategy's Sharpe Ratio.  For clarification, "original" in this context refers to the strategies that were pre-selected during the global training period, as described in section Selected Combinations. This method differs from the standard approach by using random EMA parameters drawn from a predefined set described in the section [Exponential Moving Average Crossover], whereas the standard approach selects the best EMA parameters identified during the training period for each walk-forward step.
This allows assessing the performance of the selection method compared to random selection. A bootstrap procedure was performed on the global training period. The cost assumption remains the same as in previous calculations. The global training data period covers data from February 8, 2018, to September 1, 2019. Each transaction incurs a 0.1\% cost. Changing positions from short to long requires two transactions, resulting in a total cost of 0.2\% }
\end{tablenotes}
\end{threeparttable}
\end{table}

Table \ref{desc_boot_ma_stats} presents the descriptive statistics for
Sharpe Ratio achieved during bootstrap iterations.

Key findings:

\begin{itemize}
\tightlist
\item
  The average Sharpe Ratios across both bootstrap strategies are very
  close
\item
  The maximum Sharpe Ratio is 40-50\% higher than the original
  strategies Sharpe Ratios
\item
  The original strategies' Sharpe Ratios fall within 1 standard deviation
  of the average Sharpe Ratio across bootstrap iterations. For
  clarification, ``original'' in this context refers to the strategies
  that were pre-selected during the global training period, as described in section Selected Combinations.
\end{itemize}

\begin{table}[H]

\caption{\label{tab:boot_ma_stats}\label{desc_boot_ma_stats} The Descriptive Statistics of Sharpe Ratios  - Bootstrap Evaluation of Subsampled EMA Combinations for Walk-Forward Optimization}
\centering
\fontsize{10}{12}\selectfont
\begin{threeparttable}
\begin{tabular}{rrrrrrrrr}
\toprule
Description & Mean & SD & Median & Min & Max & Range & Skew & Kurtosis\\
\midrule
\cellcolor{gray!6}{BTC Train 7 Test 28} & \cellcolor{gray!6}{0.7847} & \cellcolor{gray!6}{0.3259} & \cellcolor{gray!6}{0.7888} & \cellcolor{gray!6}{-0.4584} & \cellcolor{gray!6}{1.777} & \cellcolor{gray!6}{2.236} & \cellcolor{gray!6}{-0.0856} & \cellcolor{gray!6}{0.1486}\\
BTC Train 14 Test 10 & 0.7901 & 0.3448 & 0.7948 & -0.3549 & 1.716 & 2.071 & -0.1403 & -0.1138\\
\bottomrule
\end{tabular}
\begin{tablenotes}[para]
\small
\item \textit{\tiny{ Note: }} 
\item \tiny{ This method differs from the standard approach by using random EMA parameters drawn from a predefined set described in the section [Exponential Moving Average Crossover], 
whereas the standard approach selects the best EMA parameters identified during the training period for each walk-forward step.
This allows assessing the performance of the selection method compared to random selection. A bootstrap procedure was performed on the global training period. The cost assumption remains the same as in previous calculations. The global training data period covers data from February 8, 2018, to September 1, 2019. Each transaction incurs a 0.1\% cost. Changing positions from short to long requires two transactions, resulting in a total cost of 0.2\% }
\end{tablenotes}
\end{threeparttable}
\end{table}

\hypertarget{results---bootstrap-evaluation-of-shuffled-transaction-blocks}{%
\paragraph{Results—Bootstrap Evaluation of Shuffled Transaction
blocks}\label{results---bootstrap-evaluation-of-shuffled-transaction-blocks}}

Both strategy combinations selected in the
\protect\hyperlink{selected-combinations}{Selected Combinations} section
were evaluated for statistical significance using the method detailed in
the {[}Bootstrap Evaluation of Random EMA Combinations in Walk- Forward
Optimization{]} section. This method of bootstrap shuffles blocks of
original transactional positions generated by the walk-forward within the
global training period.

Key findings:

\begin{itemize}
\tightlist
\item
  Both combinations were statistically significant
\item
  5\% statistical significance was reached by both sets of strategies.
\end{itemize}

\begin{table}[H]

\caption{\label{tab:table_boot_position }Results - Bootstrap Evaluation of Shuffled Transaction Blocks}
\centering
\fontsize{10}{12}\selectfont
\begin{threeparttable}
\begin{tabular}{rrrrr}
\toprule
Description & \makecell[r]{Original\\Sharpe Ratio} & \makecell[r]{Total No. of\\bootstrap iterations} & \makecell[r]{No. Bootstrap Iterations\\with Higher Sharpe Ratio} & Significance\%\\
\midrule
\cellcolor{gray!6}{BTC Train 7 Test 28} & \cellcolor{gray!6}{1.252} & \cellcolor{gray!6}{1000} & \cellcolor{gray!6}{35} & \cellcolor{gray!6}{3.5}\\
BTC Train 14 Test 10 & 1.178 & 1000 & 44 & 4.4\\
\bottomrule
\end{tabular}
\begin{tablenotes}[para]
\small
\item \textit{\tiny{ Note: }} 
\item \tiny{ Original Sharpe Ratio - Sharpe Ratios of the Pre-Selected Best Strategies from the Global Training Period,  Total No. of bootstrap iterations - Number of times the bootstrap procedure was repeated, No. Bootstrap Iterations with Higher Sharpe Ratio - Number of bootstrap iterations that achieved a Sharpe Ratio exceeding the original strategy's Sharpe Ratio.  For clarification, "original" in this context refers to the strategies that were pre-selected during the global training period, as described in section Selected Combinations. Bootstrap Evaluation of Shuffled Transaction Blocks shuffles blocks of original transactional positions generated by the walk-forward within the global training period. The global training data period covers data from February 8, 2018, to September 1, 2019. Each transaction incurs a 0.1\% cost. Changing positions from short to long requires two transactions, resulting in a total cost of 0.2\% }
\end{tablenotes}
\end{threeparttable}
\end{table}

Table \ref{desc_boot_pos_stats} presents the descriptive statistics of the Sharpe Ratios for random strategies generated through bootstrapping.

Key findings :

\begin{itemize}
\tightlist
\item
  The maximum Sharpe Ratio achieved here is significantly higher than
  that reached by the original strategy. This suggests there could be a
  much better method for selecting the EMA for the walk-forward testing
  period than simply selecting the combination that performs best in the
  walk-forward training period.
\item
  The bootstrap based on shuffled positions from the 7/28 strategy has a
  smaller range and maximum value compared to the 14/10 strategy.
\item
  The average Sharpe Ratio across bootstrap iterations falls below zero
  for both strategies.
\item
  Standard Deviation is much higher than in the other bootstrap method
\item
  The original strategies' Sharpe Ratios fall within 1 standard deviation of the average Sharpe Ratio across bootstrap iterations. For
  clarification, ``original'' in this context refers to the strategies
  that were pre-selected during the global training period, as described in section Selected Combinations.
\end{itemize}

\begin{table}[H]

\caption{\label{tab:boot_positions_stats}\label{desc_boot_pos_stats} Descriptive Statistics of Sharpe Ratios  - Bootstrap Evaluation of Shuffled Transaction blocks}
\centering
\fontsize{10}{12}\selectfont
\begin{threeparttable}
\begin{tabular}{rrrrrrrrr}
\toprule
Description & Mean & SD & Median & Min & Max & Range & Skew & Kurtosis\\
\midrule
\cellcolor{gray!6}{BTC Train 7 Test 28} & \cellcolor{gray!6}{-0.2373} & \cellcolor{gray!6}{0.8083} & \cellcolor{gray!6}{-0.2391} & \cellcolor{gray!6}{-2.789} & \cellcolor{gray!6}{2.273} & \cellcolor{gray!6}{5.062} & \cellcolor{gray!6}{-0.0440} & \cellcolor{gray!6}{-0.0346}\\
BTC Train 14 Test 10 & -0.1737 & 0.8012 & -0.1955 & -3.098 & 2.721 & 5.819 & -0.0435 & 0.1718\\
\bottomrule
\end{tabular}
\begin{tablenotes}[para]
\small
\item \textit{\tiny{ Note: }} 
\item \tiny{ Bootstrap Evaluation of Shuffled Transaction blocks shuffles blocks of original transactional positions generated by the walk-forward within the global training period. The global training data period covers data from February 8, 2018, to September 1, 2019. Each transaction incurs a 0.1\% cost. Changing positions from short to long requires two transactions, resulting in a total cost of 0.2\% }
\end{tablenotes}
\end{threeparttable}
\end{table}

\hypertarget{cost-sensitivity-1}{%
\subsection{Cost Sensitivity}\label{cost-sensitivity-1}}

Recognizing the significant influence of cost on strategy outcomes, a cost sensitivity analysis investigated how the result characteristics change
with varying cost levels, isolating cost as the sole variable. 

\hypertarget{strategy-metrics---cost-sensitivity---global-training-data-period}{%
\subsubsection{Strategy Metrics - Cost sensitivity - Global Training
Data
Period}\label{strategy-metrics---cost-sensitivity---global-training-data-period}}

Key findings:

\begin{itemize}
\tightlist
\item
  Profitability: Increasing transaction cost per trade from 0.1\% to
  0.2\% leads to a decrease in annualized profit, while volatility
  remains largely unchanged.
\item
  Sensitivity to Cost Increases: Further increases of 0.1\% per
  transaction (reaching 0.3\%) result in an average 0.3\% drop in
  Annualized Mean Return.
\item
  Drawdown: Maximum drawdown is affected heavily by rising transaction
  costs.
\item
  Information Ratio (IR): IR decreases non-linearly with rising cost.
\item
  Profitability Threshold: The strategy becomes unprofitable at a cost
  level of around 0.4\% per transaction.
\end{itemize}

\begin{table}[H]
\caption{\label{tab:trading_stats_cost_sense}Cost sensitivity}
\centering
\fontsize{10}{12}\selectfont

\resizebox{\textwidth}{!}{%
\begin{tabular}{lrrrrrr}
\toprule
Cost level & \makecell[r]{Annualized\\Mean Return} & \makecell[r]{Annualized\\Volatility} & Sharpe Ratio & Information Ratio** & Max Drawdown & Sortino Ratio\\
\midrule
\cellcolor{gray!6}{0.05\%} & \cellcolor{gray!6}{1.1068} & \cellcolor{gray!6}{0.7569} & \cellcolor{gray!6}{1.4623} & \cellcolor{gray!6}{6.6457} & \cellcolor{gray!6}{0.3408} & \cellcolor{gray!6}{2.1035}\\
0.07\% & 1.0419 & 0.7570 & 1.3763 & 5.7491 & 0.3442 & 1.9788\\
\cellcolor{gray!6}{0.10\%} & \cellcolor{gray!6}{0.9483} & \cellcolor{gray!6}{0.7572} & \cellcolor{gray!6}{1.2524} & \cellcolor{gray!6}{4.6218} & \cellcolor{gray!6}{0.3516} & \cellcolor{gray!6}{1.7990}\\
0.20\% & 0.6197 & 0.7584 & 0.8171 & 1.5867 & 0.4240 & 1.1699\\
\cellcolor{gray!6}{0.30\%} & \cellcolor{gray!6}{0.2950} & \cellcolor{gray!6}{0.7604} & \cellcolor{gray!6}{0.3879} & \cellcolor{gray!6}{0.2902} & \cellcolor{gray!6}{0.4979} & \cellcolor{gray!6}{0.5531}\\
\addlinespace
0.40\% & -0.0297 & 0.7633 & -0.0390 & -0.0024 & 0.6448 & -0.0553\\
\cellcolor{gray!6}{0.50\%} & \cellcolor{gray!6}{-0.3545} & \cellcolor{gray!6}{0.7670} & \cellcolor{gray!6}{-0.4622} & \cellcolor{gray!6}{-0.2193} & \cellcolor{gray!6}{0.7788} & \cellcolor{gray!6}{-0.6519}\\
\bottomrule
\end{tabular}%
}
\begin{threeparttable}
\TPTnoteSettings{}
\begin{tablenotes}
\footnotesize
\item \begin{minipage}{\textwidth}
\textit{Note:} Strategy metrics for the Exponential Moving Average strategy optimized with a walk-forward process \\ (7 days training, 28 days testing) applied in the global training period, assuming different transaction costs.
\end{minipage}
\end{tablenotes}
\end{threeparttable}
\end{table}

\begin{figure}[H]
\includegraphics{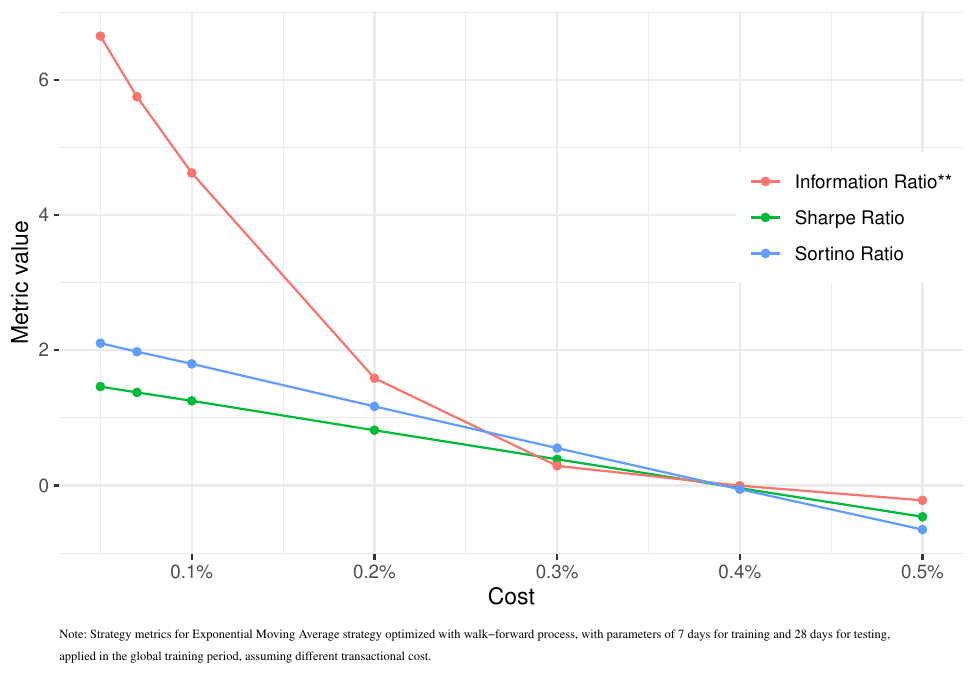} \caption{\label{cost_sense_metric}Strategy metrics for different level of transactional cost}\label{fig:coste_sense_by_cost}
\end{figure}

Strategy Metrics Visualization - Cost sensitivity - Global Training Data Period (Figure \ref{fig:coste_sense_by_cost})

To visualize the impact of transaction costs on the strategy's
performance, metrics are presented graphically. The horizontal axis
represents the level of transaction cost, while the vertical axis
displays unitless metric values.

Key findings:

\begin{itemize}
\tightlist
\item
  Information Ratio** appears to be the metric most significantly
  affected by cost changes.
\item
  The decrease in Information Ratio** is non-linear, with a steeper
  decline observed at the beginning of the cost range compared to the
  end.
\item
  Other metrics exhibit a nearly linear decrease, with some fluctuations
  at the start of the cost range.
\item
  Based on the observed trends, the breakeven point, where the strategy's performance breakeven transaction costs, can be estimated
  to be around 0.36\% cost per transaction.
\end{itemize}

\begin{figure}[H]
\includegraphics{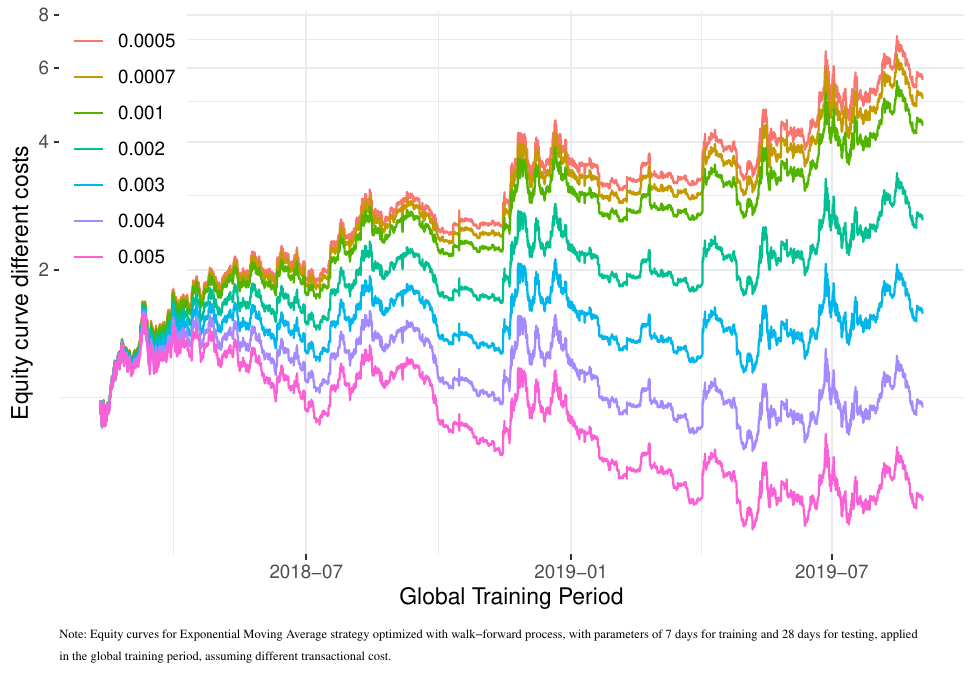} \caption{\label{cost_sense_equity_curves} Equity curves for different level of transactional cost}\label{fig:equity_curve_cost_sense}
\end{figure}

Equity Curves - Cost sensitivity - Global Training Data Period (Figure \ref{fig:equity_curve_cost_sense})

Key findings:

\begin{itemize}
\tightlist
\item
  Increasing cost changes the direction of the original equity curve to downward sloping
\item
  Equity curves look the same, and their volatility did not increase just
  change direction
\item
  The first equity curve, which is unprofitable, is for a cost of 0.4\%, the breakeven point is somewhere between 0.3\% to 0.4\%
\end{itemize}

\hypertarget{results---unseen-data-period}{%
\section{Results - Unseen Data
Period}\label{results---unseen-data-period}}

Following the optimization within the global training period, the two
walk-forward parameter combinations yielding the highest Sharpe ratios
were evaluated on unseen data. To further validate the method's
robustness, the strategy was then applied to two other cryptocurrencies, Ethereum (ETH) and Binance Coin (BNB), which were not
included in the global training period optimization. The selection of
these additional assets was driven by their market capitalization. To
mitigate the potential for cherry-picking bias, the strategy was
evaluated on the unseen data period only once. This approach ensured
that the results and parameters were not subjected to any manipulation
or modification after the initial evaluation.

\hypertarget{results-for-single-cryptocurrencies-on-unseen-data}{%
\subsection{Results for single cryptocurrencies on unseen
data}\label{results-for-single-cryptocurrencies-on-unseen-data}}

Table \ref{trad_stats_unseen} shows metrics of trading strategies
executed on the unseen data period. The walk-forward parameters of
strategies were derived from the global training period. Combinations of
parameters that achieved the highest Sharpe ratio were selected. Table
\ref{trad_stats_unseen} also includes a benchmark, which is Buy-and-Hold
strategy.

The best walk-forward window size selected from the Global Training
Data Period:

\begin{itemize}
\tightlist
\item
  Train 7 days test 28 days (WF 7/28)
\item
  Train 14 days test 10 days (WF 14/10)
\end{itemize}

Key findings:

\begin{itemize}
\tightlist
\item
  Buy-and-Hold - Achieved the highest Sharpe Ratio for all cryptocurrencies, outperforming the walk-forward strategies for the given cryptocurrency Buy-and-Hold for ETH and BNB had higher Sharpe Ratios than BTC.
\item
  Selected strategy performance - WF 14/10 outperformed the WF 7/28
  strategy across all cryptocurrencies, with an average improvement of
  10\% in Sharpe Ratio for ETH and BNB, and a 40\% higher Sharpe Ratio for BTC. Notably, both ETH and BNB were not included in the
  optimization process in the global training period, suggesting
  potential generalizability of the strategy.
\item
  Comparison to Buy-and-Hold - No strategy surpassed the respective
  asset's Buy-and-Hold performance in terms of Sharpe ratio. For BTC,
  the WF 14/10 strategy came closest, with Buy-and-Hold outperforming by
  less than 2\%. In contrast, ETH and BNB experienced a more significant
  performance gap, with Buy-and-Hold outperforming the WF 14/10 strategy
  by 20\% and 15\%, respectively. The WF 14/10 strategy outperformed
  Buy-and-Hold on Information Ratio** for ETH and by around 30\% for
  BTC. For BTC only, the WF 14/10 strategy surpassed the performance of
  Sortino Buy-and-Hold.
\item
  Performance by risk-adjusted return measure - (Sharpe Ratio, Sortino
  Ratio, Information Ratio**) Buy-and-Hold strategies for ETH and BNB
  achieved the highest Sharpe Ratios, followed by the WF 14/10 strategy
  in third place. BTC WF 14/10 strategy outperformed Buy-and-Hold in
  terms of the Sortino ratio, indicating better risk-adjusted returns. ETH
  and BNB - WF 14/10 strategy underperformed Buy-and-Hold by a smaller
  margin compared to the Sharpe ratio metric, with Buy-and-Hold
  outperforming by only 13\% and 7\% for ETH and BNB, respectively. In terms of Information Ratio**, Buy-and-Hold significantly outperformed
  the WF 14/10 by 35\%, while the WF 14/10 reached the highest Information
  Ratio** among all strategies.
\end{itemize}

The overall observations:

\begin{itemize}
\tightlist
\item
  Buy-and-Hold remains a strong benchmark for all three
  cryptocurrencies, particularly in terms of Sharpe ratio.
\item
  The WF 14/10 strategy shows promising results across different asset classes, exceeding other tested strategies and exhibiting potential
  generalizability.
\item
  The choice of risk-adjusted return measure (Sharpe vs.~Sortino) can
  influence the interpretation of performance, particularly when
  comparing strategies with different risk profiles.
\end{itemize}

\begin{table}[H]

\caption{\label{tab:trading_stats_unseen }\label{trad_stats_unseen} Performance Metrics -  Unseen Data Period -  Walk-Forward Top Combinations from Global Training Data Period}
\centering
\fontsize{10}{12}\selectfont
\begin{threeparttable}
\begin{tabular}{rrrrrrr}
\toprule
Description & \makecell[r]{Annualized\\Mean\\Return} & \makecell[r]{Annualized\\Volatility} & \makecell[r]{Sharpe\\Ratio} & \makecell[r]{Information\\Ratio**} & \makecell[r]{Max\\Drawdown} & \makecell[r]{Sortino\\Ratio}\\
\midrule
\cellcolor{gray!6}{BTC Buy and Hold} & \cellcolor{gray!6}{\textbf{0.9267}} & \cellcolor{gray!6}{\underline{\textbf{0.8214}}} & \cellcolor{gray!6}{\textbf{1.1281}} & \cellcolor{gray!6}{2.0485} & \cellcolor{gray!6}{0.6228} & \cellcolor{gray!6}{1.554}\\
BTC Train 7 Test 28 & 0.6828 & 0.8219 & 0.8307 & 1.3191 & 0.5849 & 1.2198\\
\cellcolor{gray!6}{BTC Train 14 Test 10} & \cellcolor{gray!6}{0.9091} & \cellcolor{gray!6}{0.8216} & \cellcolor{gray!6}{1.1064} & \cellcolor{gray!6}{\textbf{2.6275}} & \cellcolor{gray!6}{\underline{\textbf{0.5243}}} & \cellcolor{gray!6}{\textbf{1.6249}}\\
\addlinespace
ETH Buy and Hold & \textbf{1.5771} & \textbf{1.0264} & \underline{\textbf{1.5365}} & 4.0574 & 0.6937 & \underline{\textbf{2.1089}}\\
\cellcolor{gray!6}{ETH Train 7 Test 28} & \cellcolor{gray!6}{1.2408} & \cellcolor{gray!6}{1.0268} & \cellcolor{gray!6}{1.2085} & \cellcolor{gray!6}{2.0651} & \cellcolor{gray!6}{0.7696} & \cellcolor{gray!6}{1.7669}\\
ETH Train 14 Test 10 & 1.3727 & 1.0266 & 1.3371 & \underline{\textbf{4.2635}} & \textbf{0.6177} & 1.9519\\
\addlinespace
\cellcolor{gray!6}{BNB Buy and Hold} & \cellcolor{gray!6}{\underline{\textbf{1.7142}}} & \cellcolor{gray!6}{\textbf{1.1705}} & \cellcolor{gray!6}{\textbf{1.4644}} & \cellcolor{gray!6}{\textbf{3.5801}} & \cellcolor{gray!6}{0.7483} & \cellcolor{gray!6}{\textbf{2.0211}}\\
BNB Train 7 Test 28 & 1.2879 & 1.1712 & 1.0997 & 1.8793 & 0.7798 & 1.6298\\
\cellcolor{gray!6}{BNB Train 14 Test 10} & \cellcolor{gray!6}{1.403} & \cellcolor{gray!6}{1.171} & \cellcolor{gray!6}{1.1982} & \cellcolor{gray!6}{2.6565} & \cellcolor{gray!6}{\textbf{0.7275}} & \cellcolor{gray!6}{1.785}\\
\bottomrule
\end{tabular}
\begin{tablenotes}[para]
\small
\item \textit{\tiny{ Note: }} 
\item \tiny{ Performance metrics for the combination of strategies with the highest Sharpe Ratio identified on the global training period and applied to the unseen period. The global training period was used in the case of BTC to find optimal walk-forward parameters and train/test day lengths of 7/28 and 14/10. 
Strategy with these combinations of parameters executed on the unseen data period using BTC and the cryptocurrencies not used before i.e., ETH and BNB. The bolded values indicate the best metric within three strategies for each cryptocurrency separately, while the underlined values indicate the best metrics globally for all nine strategies. The unseen data period covers data from November 7, 2019, to August 22, 2021.   Each transaction incurs a 0.1\% cost. Changing positions from short to long requires two transactions, resulting in a total cost of 0.2\% }
\end{tablenotes}
\end{threeparttable}
\end{table}

\begin{figure}[H]
\includegraphics{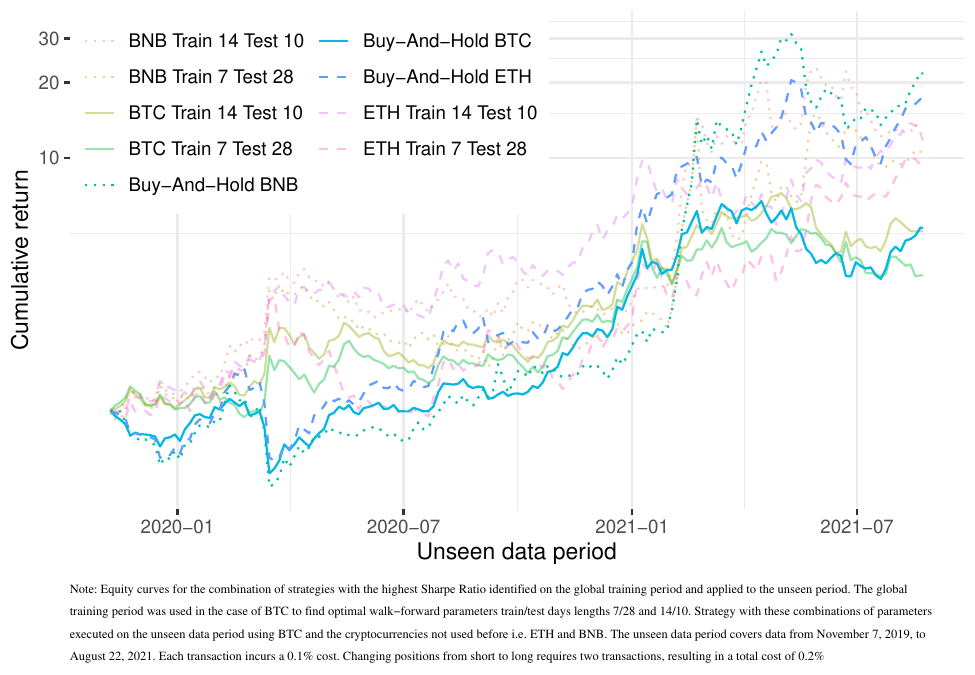} \caption{\label{trad_stats_unseen_chart} Equity Curves -  Unseen Data Period -  Walk-Forward Top Combinations from Global Training Data Period}\label{fig:equity_curve_unseen_data}
\end{figure}

Equity Curves - Unseen Data Period - Walk-Forward Top Combinations from
the Global Training Data Period (Figure \ref{fig:equity_curve_unseen_data})

Key findings:

\begin{itemize}
\tightlist
\item
  BNB Buy-and-Hold finished with the highest cumulative gains of all strategies, underperforming in 2020 and overperforming in 2021.
\item
  Both variations of strategy performed better than Buy-and-Hold in 2020; BNB WF 14/10 was unbeatable in 2020.
\item
  The situation changed in 2021 when Buy-and-Hold take the lead and ETH, together with BNB, recovered all it loses.
\item
  BTC strategies behaved very similarly to Buy-and-Hold. Buy-and-Hold
  underperformed throughout the entire period, only delivering some
  gains at the end of the unseen period. This resulted in a walk-forward
  strategies slightly underperforming Buy-and-Hold in terms of Sharpe
  Ratio
\item
  All variations of strategies provided good protection against losses in 2020; when Buy-and-Hold lost significantly in March 2020, strategies
  observed gains.
\item
  Problematic periods when the strategy started to lose steadily were the periods when Buy-and-Hold recovered and flat periods with low
  volatility.
\item
  Interestingly, walk-forward strategies dominated throughout 2020,
  outperforming their Buy-and-Hold counterparts across all
  cryptocurrencies. If the study had concluded at the end of 2020,
  walk-forward strategies would likely have been the clear winner in
  every category. However, 2021 saw a shift, with walk-forward
  strategies underperforming against Buy-and-Hold strategies. The reason
  for this reversal remains unclear. However, it's possible that
  walk-forward strategies based on parameters that were optimal in 2018
  and 2019 are no longer providing an edge. Therefore, walk-forward
  strategies may benefit from retraining and recalibration to identify
  new optimal windows in changing market conditions.
\end{itemize}

\hypertarget{results-for-portfolio-of-cryptocurrencies-on-unseen-data}{%
\subsection{Results for portfolio of cryptocurrencies on unseen
data}\label{results-for-portfolio-of-cryptocurrencies-on-unseen-data}}

Three portfolios were tested on the Unseen Data Period to evaluate their
performance:

\begin{itemize}
\tightlist
\item
  Buy-and-Hold: Benchmark portfolio holding Bitcoin (BTC), Ethereum (ETH), Binance Coin (BNB) no rebalancing during the whole period.
\item
  Portfolio 1 (the walk-forward training 7 days/testing 28 days - WF 7/28): Strategy using an Exponential Moving Averages and the walk-forward with a 7-day training period and 28-day testing period, using 3 different cryptocurrencies: Bitcoin (BTC), Ethereum (ETH), Binance Coin (BNB). Equal weights were assigned initially, with no rebalancing during the whole period.
\item
  Portfolio 2 (the walk-forward training 14 days/testing 10 days- WF 14/10): Similar strategy to Portfolio 1, but with a 14-day training period and 10-day testing period, applied to Bitcoin (BTC), Ethereum (ETH), and Binance Coin (BNB). Equal weights were assigned initially, with no rebalancing during the whole period.
\item
  All portfolios combined: Buy-and-Hold, Portfolio 1, Portfolio 2
  combined to assess the potential performance benefits of combining
  active and passive strategies.
\end{itemize}

Performance Metrics - Portfolio - Unseen Data Period

Key finding:

\begin{itemize}
\item
  Sharpe Ratio: All portfolios combined reached the highest Sharpe
  Ratio, beating the second Buy-and-Hold portfolio by 20\%. Combining all
  portfolios decreased volatility significantly.
\item
  Information Ratio (IR): The combined portfolio strategy achieved the
  highest performance once again, surpassing second-placed WF 14/10 by a
  significant margin and doubling its value.
\item
  Sortino Ratio: All portfolios combined scored the highest Sortino Ratio,
  outperforming the second WF 14/10, and the Buy-and-Hold portfolio by more than
  30\%.
\item
  Maximum Drawdown: Combining all portfolios also resulted in the
  smallest maximum drawdown. Maximum drawdown decreased from nearly 70\%
  for Buy-and-Hold to around 45\%. This is particularly outstanding
  considering that walk-forward strategies experienced an average
  drawdown of around 60\%.
\item
  Combining all the portfolios significantly improved each
  performance metric. This resulted in decreased volatility and maximum
  drawdown and unleashed the power of diversification.
\end{itemize}

\begin{table}[H]

\caption{\label{tab:trading_stats_unseen}\label{trad_stats_portfolio_unseen} Performance Metrics -  Unseen Data Period -  Walk-Forward Top Combinations from Global Training Data Period - Portfolio}
\centering
\fontsize{10}{12}\selectfont
\begin{threeparttable}
\resizebox{\textwidth}{!}{%
\begin{tabular}{rrrrrrr}
\toprule
Description & \makecell[r]{Annualized\\Mean\\Return} & \makecell[r]{Annualized\\Volatility} & \makecell[r]{Sharpe\\Ratio} & \makecell[r]{Information\\Ratio**} & \makecell[r]{Max\\Drawdown} & \makecell[r]{Sortino\\Ratio}\\
\midrule
\cellcolor{gray!6}{Portfolio 3 currencies Train 7 Test 28} & \cellcolor{gray!6}{1.129} & \cellcolor{gray!6}{0.889} & \cellcolor{gray!6}{1.27} & \cellcolor{gray!6}{3.047} & \cellcolor{gray!6}{0.619} & \cellcolor{gray!6}{1.874}\\
Portfolio 3 currencies Train 14 Test 10 & 1.269 & 0.907 & 1.4 & 4.274 & 0.569 & 2.067\\
\addlinespace
\cellcolor{gray!6}{Portfolio 3 currencies Buy and Hold} & \cellcolor{gray!6}{\textbf{1.495}} & \cellcolor{gray!6}{0.97} & \cellcolor{gray!6}{1.542} & \cellcolor{gray!6}{3.87} & \cellcolor{gray!6}{0.683} & \cellcolor{gray!6}{2.075}\\
All Portfolios Combined & 1.318 & \textbf{0.686} & \textbf{1.921} & \textbf{9.369} & \textbf{0.438} & \textbf{2.757}\\
\bottomrule
\end{tabular}%
}
\begin{tablenotes}
\footnotesize
\item[]
\begin{minipage}{0.95\textwidth}{Note: Trading metrics for portfolios with equal weights at inception: 1. Buy-and-Hold portfolio - BTC, ETH, BNB equally weighted. 2. Portfolio - walk-forward strategy with training for 7 days and testing for 28 days - strategy applied to BTC, ETH, and BNB equally weighted. 3. Portfolio - walk-forward strategy with training for 14 days and testing for 10 days -  applied to BTC, ETH, and BNB equally weighted. 
No rebalancing during the whole period. The unseen data period covers data from November 7, 2019, to August 22, 2021. The bolded values indicate the best metric within three strategies.  Each transaction incurs a 0.1\% cost. Changing positions from short to long requires two transactions, resulting in a total cost of 0.2\% }
\end{minipage}
\end{tablenotes}
\end{threeparttable}
\end{table}

\begin{figure}[H]
\includegraphics{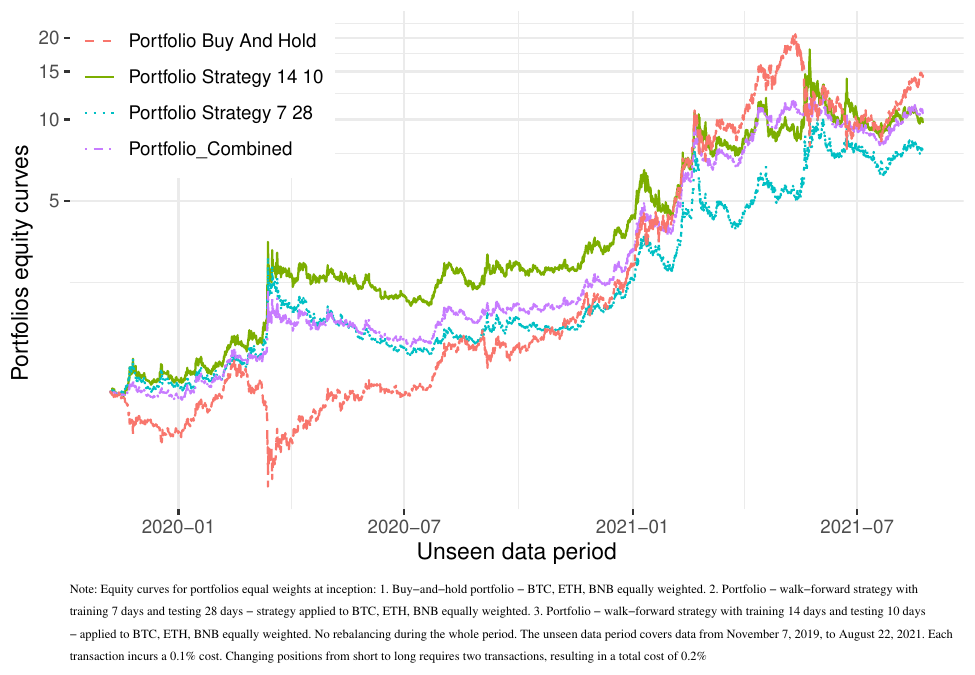} \caption{\label{portfolio_unseen_chart} Equity Curves -  Unseen Data Period -  PortfolioofCrypotcurrencies}\label{fig:equity_curve_portfolio_unseen_data}
\end{figure}

Equity curves - Portfolio - Unseen Data Period (Figure \ref{fig:equity_curve_portfolio_unseen_data})

Key Findings:

\begin{itemize}
\tightlist
\item
  Profitability: Most profits for the MA/WF strategies occurred during periods of high volatility, coinciding with significant losses of the Buy-and-Hold portfolio.
\item
  Recovery Phases: During Buy-and-Hold portfolio recovery periods, both
  WF portfolios lost momentum or declined.
\item
  Upward Trends: In stable upward markets, the WF portfolios slightly
  underperformed the Buy-and-Hold portfolio.
\item
  Recovery Behavior: Portfolio WF 7/28 lost more than Portfolio WF 14/10
  during Buy-and-Hold recovery phases.
\item
  Overall Performance: For about 70\% of the time in the Unseen Data Period, the WF portfolios' strategies outperformed the Buy-and-Hold portfolio, but ultimately, the Buy-and-Hold portfolio achieved the highest
  overall return.
\end{itemize}

\hypertarget{summary}{%
\section{Summary}\label{summary}}

We divided the data into roughly equal halves: a global training period and an unseen period, each spanning over 18 months of intraday data. Instead of optimizing EMA parameters directly, we used a novel walk-forward process to optimize training and testing window lengths. This involved multiple walk-forward optimizations on the global training period to find combinations maximizing risk-adjusted returns via the Robust Sharpe Ratio.

To avoid cherry-picking bias, unseen data was used only once without further optimizations. Two optimal walk-forward parameter sets from training were applied to the unseen period. These strategies slightly underperformed Buy-and-Hold in the Sharpe Ratio but outperformed in maximum drawdown. Results were mixed for Sortino and Information Ratios: one set outperformed Buy-and-Hold, the other lagged. Notably, the top training performer underperformed in the unseen period.

Strategy robustness was validated by applying the same parameters (optimized on Bitcoin) to Ethereum and Binance Coin in the unseen period—assets unused during optimization. Results mirrored Bitcoin: Buy-and-Hold outperformed both walk-forward strategies more in the Sharpe Ratio; for other metrics, outcomes were mixed. The 14/10-day training/testing strategy beat Buy-and-Hold in Information Ratio and maximum drawdown for ETH, but for BNB, Buy-and-Hold excelled in most metrics except drawdown.

Overall, the top training parameters (7/28 days) underperformed Buy-and-Hold and the 14/10-day alternative in the unseen period, questioning the parameter selection method's optimality.

Sharpe Ratio performance dropped 20\% on average from training (top 2 of 81 combinations) to testing; the best combination fell 30\%, the second 10\%.

From unseen period results across three cryptocurrencies, we built portfolios using: (1) walk-forward 7/28 days, (2) walk-forward 14/10 days, and (3) Buy-and-Hold. Both walk-forward portfolios showed lower drawdown (7/28: 10\% less; 14/10: 20\% less) and volatility than Buy-and-Hold. Buy-and-Hold led in Sharpe Ratio, but the 14/10 walk-forward beat it in Information Ratio, while 7/28 lagged 20\%. Equity curves revealed Buy-and-Hold gains surged in the unseen period's second half, with walk-forward leading initially.

Findings show the EMA strategy excelled in high-volatility periods with large asset gains/losses, surpassing Buy-and-Hold, but underperformed in low-volatility phases. It consistently reduced maximum drawdown versus Buy-and-Hold.

Heuristically, equity curves indicate our strategy outperformed Buy-and-Hold over a 14-month unseen period. Under a common 2:1 training/testing ratio ending after a year, it would still excel. This prompts questions on parameter validity, duration, and retraining timing. All 60-minute frequency training combinations outperformed Buy-and-Hold, with Sharpe Ratios above 0 versus Buy-and-Hold's.

The hybrid portfolio—combining Buy-and-Hold with two walk-forward portfolios—outperformed all individuals, single assets, and portfolios across metrics, highlighting diversification value even without rebalancing.

Finally, against random strategies via two custom bootstrap methods, one showed statistical significance for both top parameter combinations.

\hypertarget{conclusions}{%
\section{Conclusions}\label{conclusions}}

The development of trading strategies in stock and cryptocurrency markets has been extensively explored through technical analysis in prior research. However, these studies frequently overlook critical elements such as transaction fees, slippage, and other trading costs. Additionally, a common but often unreported issue involves repeated optimizations on data designated as out-of-sample or the complete omission of an out-of-sample evaluation phase. Such practices foster unrealistic performance expectations, leading to substantial underperformance when strategies are implemented in live trading environments.

This study aimed to develop a trading strategy combining the Exponential Moving Average (EMA) and walk-forward optimization, assessing its performance on nearly four years of intraday cryptocurrency data. Rather than maximizing results through EMA parameter optimization, a unique approach tested how data length periods influence strategy performance. Results show that specific training/testing period lengths in walk-forward optimization yield more favorable outcomes, contradicting the Efficient Market Hypothesis, which implies no differences across period lengths.

The study addressed the following questions:  

RH1: What is the optimal length of training/testing periods for the walk-forward technique and moving average crossover strategy?  
The investigation sought ideal training and testing lengths. Initial results favored 7-day training and 28-day testing, but unseen data showed a 14-day training and 10-day testing combination as superior. Periods longer than 7 days for both may generally be advantageous.  

RH2: How do the choice of data frequency and costs impact strategy results?  
Performance was evaluated across timeframes from 1 to 60 minutes. With 0.1\% transaction costs, only intervals over 30 minutes appeared profitable, as shorter ones fail to overcome costs. Below 60 minutes should be avoided at this cost level. Breakeven was around 0.4\% per transaction for 60-minute timeframes.  

RH3: Is the tested strategy better than random?  
The strategy was compared to random approaches via two bootstrapping methods. Block bootstrap indicated superiority at 5\% confidence. Random EMA bootstrap found only one random strategy slightly better, affirming merit.  

Walk-forward optimization prevented overfitting in parameter estimation. The study examined varying training and testing lengths' effects, enhancing efficacy and theoretical understanding. Data were partitioned into global training for optimization and unseen periods for execution, minimizing bias—unlike iterative out-of-sample testing. Intraday data for Bitcoin, Ethereum, and Binance Coin were used, with Bitcoin parameters applied across assets. Limitations include order book impact (average bid-ask prices may not reflect large-order executions), bid-ask uncertainty (0.1\% cost optimistic for illiquid exchanges), holding costs (for futures), shorting availability (assumed but variable), data splitting effects, and unexplored biases.

The novelty lies in examining walk-forward window lengths, revealing performance impacts beyond arbitrary values, and opening theoretical frameworks. Findings advise dynamic period lengths and single out-of-sample evaluations. The approach reduced drawdown and improved risk-adjusted metrics, allowing equivalent profits with lower risk, enhancing stability, confidence, and crash resistance. EMA excelled in high volatility but lagged in low volatility; all 60-minute training combinations outperformed Buy-and-Hold (Sharpe >0).

The combination of Exponential Moving Average and walk-forward optimization merits further investigation. Avenues include temporal parameter validity, exploring re-evaluation every 2-3 months or varying lengths for improvement, backtesting with updated data, alternative EMA selection minimizing training/testing deviation for robustness, risk management beyond full position, and mean reversion with dynamic switching for recovery underperformance.

\printbibliography

\end{document}